\documentclass[useAMS,usenatbib,usegraphicx]{mn2e}
\usepackage{amssymb}
\usepackage{amsmath}
\usepackage{multirow}
\usepackage{booktabs}
\usepackage{caption}
\usepackage{subcaption}
\usepackage{fixltx2e}
 
\newcommand{\eg}{e.\,g., }
\newcommand{\ie}{i.\,e., }

\newcommand{\Ser}{\texttt{Ser}}
\newcommand{\DevExp}{\texttt{DevExp}}
\newcommand{\SerExp}{\texttt{SerExp}}
\newcommand{\catalog}{M2013}

\usepackage{url}
\usepackage{hyperref}

\title[2D Fit Catalog Simulations]{Simulations of single and two-component
galaxy decompositions for spectroscopically selected galaxies from the Sloan
Digital Sky Survey}
\author[Meert et al.]{Alan Meert,$^{1}$\thanks{E-mail:
ameert@physics.upenn.edu} 
Vinu Vikram,$^{1}$\thanks{E-mail: vvinuv@gmail.com} 
and Mariangela Bernardi$^{1}$\thanks{E-mail: bernardm@sas.upenn.edu} \\
$^{1}$Department of Physics and Astronomy, University of Pennsylvania, 
Philadelphia, PA 19104, USA\\}

\begin{document}
 \date{Accepted 2013 May 7.  Received 2013 April 29; in original form 2012 November 26}

\maketitle

\label{firstpage}

\begin{abstract}

We present the results of fitting simulations of an unbiased
sample
of SDSS galaxies utilizing the fitting routine GALFIT and analysis pipeline 
PyMorph. These simulations are used to test the two-dimensional decompositions
of SDSS galaxies.  The simulations show that single
S{\'e}rsic models of SDSS data are recovered with $\sigma_{\mathrm{mag}}
\approx 0.025$ mag and $\sigma_{\mathrm{radius}} \approx 5\%$. The global
values (half-light radius and magnitude) are equally well constrained when a
two-component model is used. Sub-components of two-component models present
more scatter. SDSS resolution is the primary
source of error in the recovery of models. We use a
simple statistical correction of the biases in fitted parameters,
providing an example using the S{\'e}rsic index.
Fitting a two-component S{\'e}rsic + Exponential model to a
single S{\'e}rsic galaxy results in a noisier, but unbiased, recovery of the
input parameters ($\sigma_{\mathrm{total mag}}
\approx 0.075$ mag and $\sigma_{\mathrm{radius}} \approx 10\%$); fitting a
single S{\'e}rsic profile to a  two-component system
results in biases of total magnitude and halflight radius of
$\approx 0.05-0.10$ mag and 5\%-10\% in radius. Using an
F-test to select the best fit model from among the single- and two-component
models is sufficient to remove this bias. We recommend fitting a
two-component model to all galaxies when attempting to
measure global parameters such as total magnitude and halflight radius.
\end{abstract}

\begin{keywords}
 galaxies: structural parameters -- galaxies: fundamental parameters --
 galaxies: catalogs -- methods: numerical -- galaxies: evolution 

\end{keywords}

\section{Introduction}
\label{sec:Intro}
Measurement of fundamental galaxy properties is an essential step
in analyzing galaxy structure, formation, and evolution. At
the most basic level, luminosity, size, and morphology are important 
properties, useful in evaluating dynamical and evolutionary models
\citep[\eg][]{Shankar}. Non-parametric methods exist to estimate luminosity,
size, and structure without imposing a functional form on a galaxy
\citep[\eg][]{Petrosian1976,
Abraham1996, blanton2001}. However,
non-parametric methods are sensitive to the depth of the image and to the PSF.
This can result in underestimating the luminosity and size of an object due to
missing flux in faint regions of the galaxy or when the true size of the galaxy
becomes small relative to the size of the angular PSF \citep{blanton2001,
blanton2003}.

Parametric methods offer a reasonable way to extrapolate galaxy light profiles
into fainter regions at the expense of introducing a potentially incorrect
functional form for the galaxy. 
Common functional forms used in parametric fitting include the $r^{1/4}$ and the
$r^{1/n}$ models originally
presented by \cite{DeVacouleurs1948} and \cite{Sersic1963b}.
Empirical study suggests that bulges and elliptical galaxies are better
described by de Vacouleurs profiles or S{\'e}rsic  profiles with S{\'e}rsic 
index $n\approx 4$. Disks and late-type spirals are best described by
exponential profiles or S{\'e}rsic  profiles with S{\'e}rsic  index $n\approx
1$ \citep{Freeman1970}.
More recent work has shown that the relationship between S{\'e}rsic index and
the photometric or kinematic components of a galaxy is more complicated.
Following \cite{Kent1985}, many
studies simultaneously fit a second component in order to better
accommodate the qualitative differences of bulges and disks in galaxies.
Additionally, \cite{Caon1993} showed that the S{\'e}rsic  profile is a better
fit to many early-type galaxies than the traditional de Vacouleurs profile.

There have been several catalogs of photometric galaxy decompositions
presented recently \citep{simard11, Kelvin2011, Lackner2012} with particular
interest on the applicability of large sets of image decompositions to
evolutionary models. However, systematic effects continue to be of concern, and
the reliability of two-component decompositions in cases of low to moderate
signal-to-noise are often viewed with some skepticism. In order to quantify the
systematics and robustness of the $\sim 7\times10^5$ fits of g, r, and i band
SDSS spectroscopic galaxies to be presented in \cite{Meert2013}, hereafter referred to
as \catalog, we generate simulations of single and two-component
galaxies, referred to as ``mocks,'' and fit them using the same PyMorph pipeline
\citep{PyMorph} used for
the photometric decompositions presented in \catalog. The \catalog\ catalog has
already been used in \cite{Bernardi2013} to study systematics in the
size-luminosity relation, in \cite{Shankar} to study size-evolution of
spheroids, and in \cite{huertas12} to study the environmental dependence of the
mass-size relation of early-type galaxies.

Following several detailed studies which have used simulations to test the 
robustness of different fitting algorithms \citep[\eg][]{Groth2002, GEMS2007,
Lackner2012}, 
the main goal of this paper is to test the robustness of PyMorph pipeline
software on SDSS galaxies. 
We use these simulations to test the effects of
increased signal-to-noise as
well as increased resolution, PSF errors, and sky determination. 
Our simulations are specifically applicable to SDSS galaxies and are 
useful for evaluating the decompositions presented in \catalog. We use unbiased
samples to estimate and correct the systematic error on recovered parameters
as well as estimate reasonable uncertainties on fit parameters. 

A description of the simulation process is presented in Section~\ref{sec:sims}.
This includes constructing a catalog of realistic galaxy parameters 
(Section~\ref{sec:sims:selection}); generating galaxy surface brightness
profiles based
on these parameters (Section~\ref{subsec:image_gen}); generating sky and noise
(Sections~\ref{subsec:background} and~\ref{subsec:noise}); and including seeing
effects in the final image. The completed simulations are run through the fitting
pipeline, and the fits are analyzed in Section~\ref{sec:tests}. We examine the
recovery of structural parameters in noise-free images
(Section~\ref{subsec:flat}) and parameter recovery in realistic observing
conditions including both neighboring sources and the effects of incorrect PSF
estimation
(Section~\ref{subsec:psf_test}). Recovery of mock galaxies is unbiased
for single S{\'e}rsic models. However, two-component mocks are biased when fitted with
single S{\'e}rsic profiles. This bias consists of an overestimate of
the
size and luminosity of the galaxy. PyMorph is further tested by inserting
mocks into real SDSS images to test the dependence on local density
(Section~\ref{subsec:real}). We examine dependence of the fits on resolution and
signal-to-noise (Section~\ref{subsec:sn_psf}). The effect of changing the fitted
cutout size (Section~\ref{sec:cut_size}) and the effect of incorrect background
estimation (Section~\ref{subsec:bkrd_offset}) are also examined. 
In Section~\ref{sec:discuss} we discuss the overall scatter in our fits and the
implications of the simulations.
Finally, in Section~\ref{sec:conclusions} we provide concluding remarks. 

We generate single-component S{\'e}rsic  galaxy models (hereafter
referred to as \Ser) and two forms of two-component galaxy models: 
one is a linear combination of de Vacouleurs and an exponential 
profile (\DevExp) and the other replaces the de Vacouleurs with a S{\'e}rsic 
profile (\SerExp). 
A good overview of the S{\'e}rsic  profile
used throughout this paper is presented in \cite{Graham2005}. 
Throughout the paper, a $\Lambda$CDM cosmology is assumed with
($h$,$\Omega_m$,$\Omega_{\Lambda}$) = (0.7,0.28,0.72) when necessary. 

\section{Creating the simulations} \label{sec:sims}
\subsection{Selecting the simulation catalog} \label{sec:sims:selection}
We create a set of mocks using fits  from the
photometric decompositions presented in \catalog. These galaxy parameters 
represent the r-band image decompositions of a complete sample of
the SDSS spectroscopic catalog containing all galaxies with spectroscopic
information in SDSS DR7 \citep{DR7}.

The sample contains galaxies
with extinction-corrected
r-band Petrosian magnitudes between 14 and 17.77. The lower limit of 17.77 mag
in the r-band is the lower limit for completeness of the SDSS Spectroscopic
Survey \citep{Strauss2002}. The galaxies are also
required to be identified by the SDSS \texttt{Photo} pipeline \citep[][]{Photo}
as a galaxy
(\texttt{Type = 3}), and the spectrum must also be identified as a galaxy
(\texttt{SpecClass = 2}). Additional cuts on the data following
\cite{Shen2003} and \cite{simard11} are applied.  Any galaxies with redshift $<$
0.005 are removed to prevent redshift contamination by peculiar velocity.
Galaxies with saturation, deblended as a PSF as indicated by
the \texttt{Photo} flags, or not included in the Legacy
survey\footnote{A list of fields in the Legacy survey
is provided at \url{http://www.sdss.org/dr7/coverage/allrunsdr7db.par}} are
also removed from the sample. In addition, following \cite{Strauss2002} and
\cite{simard11}, we apply a surface-brightness cut of $\mu_{\textrm{50, r}} <
23.0$ mag/arcsec$^2$ because there is incomplete spectroscopic target selection
beyond this
threshold. After applying all data cuts, a sample of 670,722 galaxies
remains. 
We select an unbiased sample of galaxies from the DR7 sample and use the fitted
models from PyMorph to generate the mocks used in this paper.

For each model (\Ser, \DevExp, and \SerExp), we select a representative sub-sample physically 
meaningful photometric decompositions. In order to ensure 
that the galaxies are representative of the full catalog, we examined 
the distributions of basic observational parameters of SDSS galaxies
(surface brightness, redshift, apparent Petrosian magnitude,
Petrosian half-light radius, and absolute magnitude). 

\begin{figure*}
\begin{minipage}{.33\linewidth}
\centering
\includegraphics[width=\columnwidth]{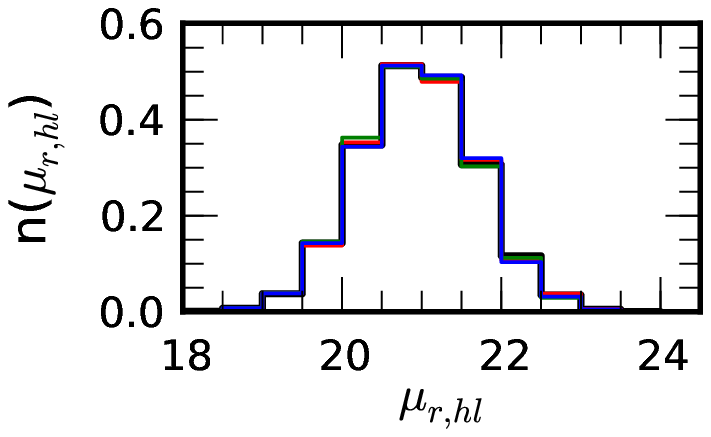}
\subcaption{}\label{subfig:orig_dist:a}
\end{minipage}
\begin{minipage}{.33\linewidth}
\centering
\includegraphics[width=\columnwidth]{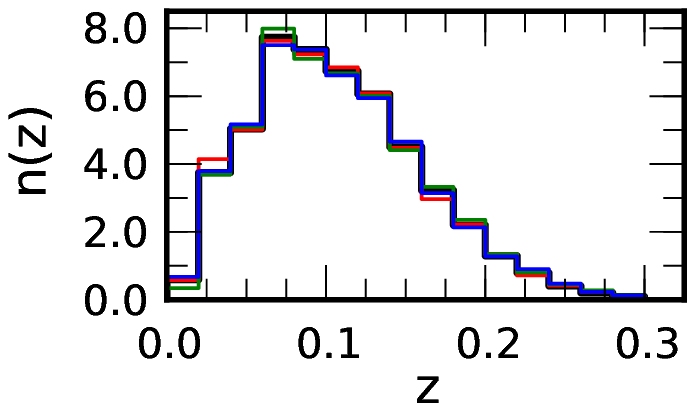}
\subcaption{}\label{subfig:orig_dist:b}
\end{minipage}
\begin{minipage}{.33\linewidth}
\centering
\includegraphics[width=\columnwidth]{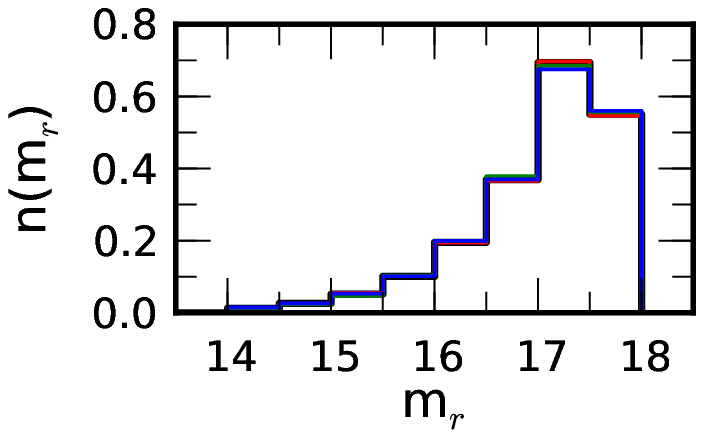}
\subcaption{}\label{subfig:orig_dist:c}
\end{minipage}

\begin{minipage}{.33\linewidth}
\centering
\includegraphics[width=\columnwidth]{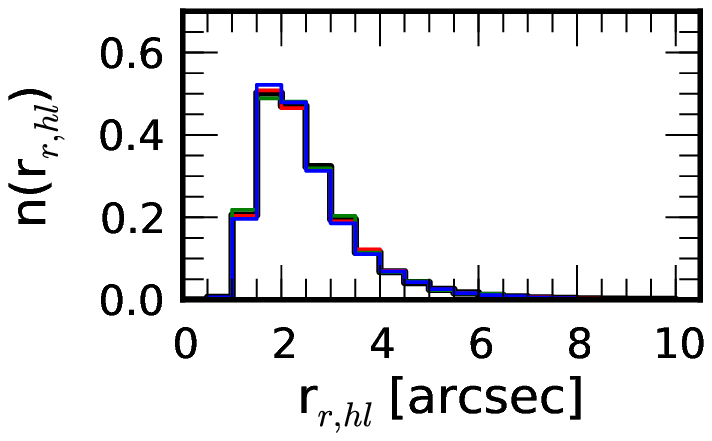}
\subcaption{}\label{subfig:orig_dist:d}
\end{minipage}
\begin{minipage}{.33\linewidth}
\centering
\includegraphics[width=\columnwidth]{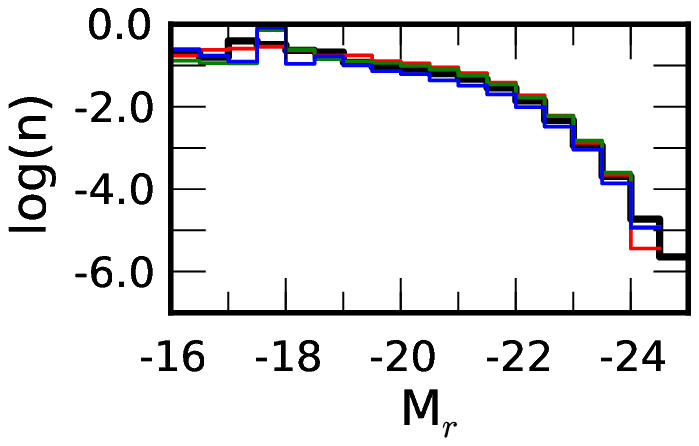}
\subcaption{}\label{subfig:orig_dist:e}
\end{minipage}
\begin{minipage}{.33\linewidth}
\centering
\includegraphics[width=\columnwidth]{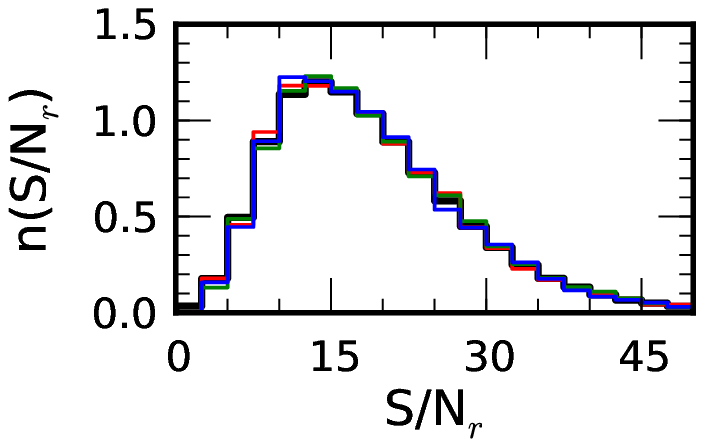}
\subcaption{}\label{subfig:orig_dist:f}
\end{minipage}

\caption{\textbf{\subref{subfig:orig_dist:a}}~The surface brightness
distribution,
\textbf{\subref{subfig:orig_dist:b}}~redshift distribution,
\textbf{\subref{subfig:orig_dist:c}}~extinction-corrected r-band Petrosian
magnitude,
\textbf{\subref{subfig:orig_dist:d}}~r-band Petrosian halflight radius,
\textbf{\subref{subfig:orig_dist:e}}~ V$_{\mathrm{max}}$-weighted
luminosity function, and \textbf{\subref{subfig:orig_dist:f}} signal-to-noise
distribution of the samples used in this paper drawn from the DR7 SDSS
spectroscopic galaxy sample. The distribution of all SDSS spectroscopic
galaxies is shown in black. Distributions of the \Ser, \DevExp, and
\SerExp\ mocks are shown in red, green, and blue, respectively. 
Bin counts are normalized to integrate to 1. The distributions of the 
mocks are representative of the full sample fitted in
\catalog\ and are appropriate to compare to the SDSS spectroscopic 
sample as verified by a Kolmogorov-Smirnov 2-sample test. The 
signal-to-noise (S/N) will be discussed further in Section~\ref{subsec:noise}. 
In calculating this S/N, we use the measurement of sky provided by 
the PyMorph pipeline rather than SDSS to identify and separate target 
counts from sky counts. PyMorph sky estimation is shown to be more 
accurate than the SDSS estimation provided in the DR7 catalog.}
\label{fig:orig_dists}
\end{figure*}

Some restrictions on fit parameters are necessary to ensure that
outliers are removed from the parameter space used to generate the
simulations. Galaxies that do not satisfy these basic cuts
are removed to ensure that the parameters used to generate the images
are physically motivated. The cuts do not significantly bias our galaxy
distribution as is shown in Figure \ref{fig:orig_dists}. The cuts are:
\begin{enumerate}
 \item Any S{\'e}rsic  components must have S{\'e}rsic  index less than 8.
\item Half-light radius of any S{\'e}rsic  component must be
less than 40 kpc.
\item In the two-component fits, the ratio of the bulge halflight radius to
disk scale radius should be less than 1, or the galaxy should be bulge
dominated (B/T $>$ 0.5).
\end{enumerate}
Conditions (i) and (ii) are used to prevent selection of \Ser\ models with
extended profiles that are likely the result of incorrect sky estimation during
the fitting process. Condition (iii) ensures that any disk dominated
galaxies have a bulge component that is smaller than the disk. 

After enforcing the cuts on the sample, 10,000 fitted galaxy profiles for each
of the \Ser, \DevExp, and \SerExp\ models are selected at random 
without regard to the morphological classification of the original galaxy. 
The fitted parameters of these sample galaxies 
are used to generate the mocks used in testing the pipeline. 

Selecting galaxy samples independent of galaxy
morphology allows the \DevExp\ and \SerExp\ samples to 
contain some galaxies that do not truly possess a second component.
Additionally, there will be some truly two-component galaxies (\ie both bulge
and disk components are present) that are misrepresented by a single S{\'e}rsic
fit. However, this sampling method will not invalidate the results of our tests.
Since we seek to test the ability to recover simulated galaxy parameters, we only
require a realistic sample of galaxy profiles. Our samples satisfy this requirement. 
Single S{\'e}rsic galaxies in the original sample, simulated as mock \Ser\
galaxies and fit with \Ser\ models, test the ability to recover S{\'e}rsic
parameters. Similarly, \Ser\ mocks with \SerExp\ models, show bias resulting 
from over-fitting a galaxy. Fitting the \SerExp\ mocks with a \Ser\ model
shows the bias due to under-fitting. 

Fitting a single-component model regardless of galaxy
structure or morphology is a common practice \citep[\eg][]{nyu_vagc,
GEMS2007, simard11}. In Figure~\ref{subfig:psf_serexp_ser} we show
that bias of 0.05 mags and 5\% of the halflight radius result from fitting a
two-component galaxy with a single component and that this bias increases to 0.1 mags
and 10\% of the halflight radius for brighter galaxies.
These biases are important in analyzing the results of a
single-component fitting catalog. For example, \cite{Bernardi2013} shows that
intermediate B/T galaxies can often be fit by S{\'e}rsic models with large
S{\'e}rsic indicies, which can lead to misclassification if cuts similar to
\cite{Shen2003} are used. 

Figure \ref{fig:orig_dists} shows the distributions of surface brightness,
redshift, extinction-corrected r-band Petrosian magnitude, 
r-band Petrosian halflight radius, and absolute magnitude of 
all SDSS spectroscopic galaxies (in black) and our simulation samples: \Ser\ 
(red), \DevExp\ (green), and \SerExp\ (blue). The distribution
of mock galaxies reproduces the observed distribution for all three samples
for each observational parameter as verified by a KS 2-sample
test. 

Figure \ref{fig:orig_dists} also presents the signal-to-noise (S/N) of the
mock samples as compared to the parent distribution. The S/N of the
images is a limiting factor in the fitting process, so care must be taken to
ensure that the S/N is not artificially increased in the simulations when
compared to true SDSS galaxies. This S/N is calculated
using the r-band Petrosian magnitude and r-band Petrosian halflight radius.
Petrosian quantities are used to make a fairer comparison among all the
samples. Because the Petrosian quantities are non-parametric, they avoid the
complications that arise in assessing the possible biases introduced during
fitting. Any differences in S/N are not large enough to significantly bias the
distributions as verified by the KS 2-sample test. Therefore, we conclude that
our samples are fair representations of the underlying distribution of SDSS
spectroscopic galaxies. The S/N is discussed further in
Section~\ref{subsec:noise}.

Testing the accuracy of the PyMorph fitting routine does not necessarily require
an unbiased parameter distribution. In reality, all that is required is a sample
with sufficient coverage of the parameter space represented by the data. The 
simulations use smooth profiles, simplifications of the true galaxies that are
observed in SDSS. Examination of the results of fitting these simplified models
and 
comparison to fits of true observed galaxies can potentially yield useful
information regarding galaxy structure. In \cite{Bernardi2013},
the simulations are used together with the decompositions of the SDSS
spectroscopic sample to characterize the scatter in the size-luminosity 
relation as well as examine possible biases. In order to make these comparisons, 
an unbiased sample is required. The distributions shown in
Figure \ref{fig:orig_dists} show that the simulations are appropriate 
to use for this purpose. 

\subsection{Generating the images} \label{subsec:image_gen}
We generate the two-dimensional normalized photon
distributions from the one-dimensional S{\'e}rsic  profiles and the
one-dimensional exponential profiles of each bulge and disk component. 
Disk components are only simulated where required, as is the case for
two-component fits. When multiple components are to be simulated, each
component's normalized photon distribution is generated separately and 
combined prior to generating the simulated exposure. 

Two-dimensional galaxy profiles are treated as azimuthally symmetric
one-dimensional
galaxy light profiles that are deformed according to an observed ellipticity.
The galaxy profiles are generated 
using the structural parameters generated from photometric decompositions
as described in the previous section. Single-component galaxy 
profiles and the bulges of two-component galaxies are generated according 
to the S{\'e}rsic  profile 
\begin{equation} \label{eq:sersic}
\begin{aligned}
& I(r)  = I_{e} \exp\left(-b_n\left[\left(\dfrac{r}{R_{e}}\right)^{\frac{1}{n}} - 1 \right]\right) \\
& b_n   = 1.9992n - 0.3271 \\
\end{aligned}
\end{equation}
where S{\'e}rsic  index (n), half-light radius ($R_{e}$), and surface brightness
at
$R_e$($I_{e}$) are selected simultaneously from the catalog described in the
previous section.

For the \DevExp\ and \SerExp\ cases, an exponential disk
(Equation~\ref{eq:sersic} with $n=1$) is added to the S{\'e}rsic  component to
model the 
disk component of the galaxies. 
The disk is modeled using a slightly modified version of
Equation~\ref{eq:sersic}.
 This model requires input parameters scale radius
($R_{d}$) and 
central surface brightness ($I_{d}$). 
\begin{equation} \label{eq:expdisk}
\begin{aligned}
& I_{Exp}(r) = I_{d} \exp\left(\dfrac{-r}{ R_{d}}\right). \\
\end{aligned}
\end{equation}

After generating the two-dimensional profile, the image is pixelated by
integrating over each pixel area. The details of this integration are
largely unimportant. However, the simulation must take careful account of the
integration in the central pixels, where the profile can vary greatly over a
single pixel. Various oversampling methods have been devised to properly
correct this common problem \citep[\eg][]{galfit, GEMS2007}. 
The simulations in this paper have been tested to ensure that the 
pixel-by-pixel integration is accurate to $\approx3\%$ of the corresponding
Poisson noise in a given pixel. Therefore, we treat the simulations as exact
calculations of the galaxy photon distributions since any noise from the
integration contributes only a small amount to the total noise budget.

The pixelated galaxy is numerically convolved with a 
PSF extracted from SDSS DR7 data using \texttt{read\_PSF} program 
distributed by SDSS\footnote{\texttt{read\_PSF} is part of the
\texttt{readAtlasImages-v5\_4\_11} package available at 
\url{http://www.sdss.org/dr7/products/images/read_psf.html}}. The choice of
this PSF is discussed in Section~\ref{subsec:psf_test}.

\subsection{Creating the background}\label{subsec:background}
Two hundred background images, each equal in size to an SDSS fpC image, are
also simulated for testing purposes. These images contain constant background and a
randomly selected field of galaxies taken from an SDSS fpC image. The
SDSS catalog provides rudimentary photometric decompositions of each star and 
galaxy. Galaxies are fit  with an exponential disk and a de Vacouleurs
($n=4$) bulge independently. The best fit is reported as a linear combination
of the two fits using the \texttt{fracdev} parameter to express the ratio of
the de Vacouleurs model to the total light in the galaxy. 

For the simulated
background used in this paper, each
galaxy is generated using the combined profile of the two fits. The
de Vacouleurs
bulge and exponential disk component are separately simulated according to the
magnitude, radius, ellipticity, and position angle reported in SDSS. 
Each component is simulated using the method
described in Section~\ref{subsec:image_gen}. The background galaxy is 
constructed by adding the two components using the \texttt{fracdev} parameter.
The galaxy is then inserted into the fpC image. Any foreground stars are also
simulated as point sources and inserted into the image.

\begin{figure} 
\includegraphics[width=\columnwidth]{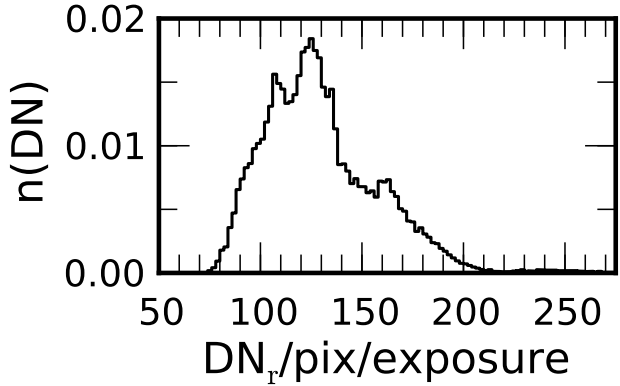}
\caption{The distribution of sky values for data in the SDSS CASJOBS catalog. 
These data are drawn from the \texttt{sky\_r} CASJOBS parameter and are 
converted into counts (DN) per pixel per standard SDSS image exposure of
54 seconds.
We use this distribution to determine the sky value used in our simulations. 
As an approximation, we use the mean value of 130 counts/pixel/exposure. }
\label{fig:sky_dist}
\end{figure}

For the background sky counts in the image, we use
the mean sky of all SDSS observations as given in the SDSS \texttt{photoobj}
table by the \texttt{sky\_r} parameter. The distribution of the sky flux is
plotted in Figure \ref{fig:sky_dist} in units of counts (or DN) per pixel per
exposure. The median and mean values for a 54 second SDSS exposure are 
$\approx$125 and $\approx$130 counts per pixel, respectively. We use 
the mean value of 130 counts per pixel as the background in our simulations.
This sky background is applied to the entire chip as a constant background; no
gradient is simulated across the image. Background gradients should be
approximately constant across a single galaxy.
This assumption is verified by inserting the simulated galaxies into real SDSS
fpC images near known clusters, where the sky contribution should be higher and
gradients are more likely. In Section~\ref{subsec:real} we show that there is
little change in the behavior of the fits in these types of environments.
 
Previous work has improved the measurements of sky background
\citep[see][]{Blanton2011}. However, these corrections tend to focus on 
areas of large, bright galaxies or on making the sky subtraction stable 
for purposes of tiling fpC images together. Since we
are only focused on maintaining the proper S/N for our simulations, the sky
levels provided in the SDSS database are sufficient, provided that they
maintain the correct S/N. We discuss the S/N distribution of our simulations
and the original SDSS galaxy sample in Section~\ref{subsec:noise} below.

Diffraction spikes and other image artifacts are not directly simulated. 
However, the SDSS \texttt{photo} pipeline often misidentifies additional
phantom sources along an observed diffraction spike. These phantom sources are
modeled in our background, and so these effects are approximately modeled. It is
reasonable to expect that the diffraction effects should
not have a large effect on the fitting process, as their elongated straight
structure does not mimic galaxy structure. The dominant effect produced by the
bright stars in the field is bias in the background estimation
in the nearby neighborhood of a star. 

After simulation of the background images, and prior to adding noise, 
each background image is convolved with a random SDSS PSF selected from
original fpC image upon which the individual image is based. 
Selecting PSFs from original SDSS images introduces a variation in PSF size
between mock galaxies inserted into images and the background galaxies.
However, this variation is not of concern for us in the fitting process because
the vast majority of galaxies (over 90\% of all galaxies) do not have neighbors
near enough to the target galaxy to require simultaneous fitting. For these
galaxies, the PSF applied to neighboring galaxies is of no interest in the
fitting process because the sources are masked out. The details of this masking 
are not discussed in the remainder of the paper. Modifying the masking conditions produce no 
noticable difference in the fitted values for our simulations.  For the remaining 10\% of
galaxies, there may be some variation in the fit due to differing PSFs. PSF 
sizes can differ between target and neighboring galaxies by up to a factor of
2. However in practice, this happens for less than 1\% of galaxies of the
galaxies with neighbors. Furthermore, incorrect
PSF tends to only cause effects at the centers of galaxies. So although using
a PSF that is different from the background PSF will affect the recovered
parameters of the neighbor, it will not affect the target galaxy. 

\subsection{Noise} \label{subsec:noise}
After generating a target galaxy and inserting it into a background,
Poisson noise is added using the average inverse gain of an SDSS image
(4.7 e$^-$/DN) and the
average contribution of the dark current and read noise, referred to as the
``dark variance,'' (1.17 DN$^2$), to determine the standard deviation for each
pixel. Specifically,  
\begin{equation}
 \label{eq:Fij}
 F_{i,j} \equiv I_{i,j} + \textrm{bkrd}_{i,j}
\end{equation}
is the total flux in pixel $(i,j)$ (\ie the sum of the source and 
background fluxes in the pixel), and 
\begin{equation}
 \label{eq:weight}
\sigma_{i,j} = \sqrt{\dfrac{F_{i,j}}{\textrm{gain}} + \textrm{dark variance}} 
\end{equation}
so 
\begin{equation}
 \left(\dfrac{\textrm{S}}{\textrm{N}}\right)_{i,j} \equiv \dfrac{I_{i,j}}
                           {\sigma_{i,j}},
\end{equation}
for a single pixel.

Since the fitting pipeline is dependent on the S/N, it is essential that the
simulated S/N is comparable to SDSS. The distribution of the average S/N per
pixel within the halflight radius for the simulations and the original galaxies
is plotted in Figure~\ref{subfig:orig_dist:f}. The S/N
distribution of simulations and the SDSS
spectroscopic galaxies agree as verified by a KS 2-sample
test, therefore the simulations appropriately approximate the S/N of SDSS
galaxies contained in \catalog.

An unbiased selection in the previously mentioned parameters is not
sufficient to guarantee fair sampling of the S/N with respect to magnitude, nor
does it prevent fictitious correlations among multiple fit parameters. In fact,
correlations among fit parameters are to be expected if the PyMorph pipeline is
robustly measuring properties of the target galaxies (many correlations exist
among physical parameters). It is difficult, and largely
unnecessary, to examine every possible relationship for correlations introduced by
biases in the sample selection process. 

Examining the S/N and the halflight radius versus apparent magnitude help
to ensure the appropriateness of the simulation. Systematic differences
in radius will lead to systematic variation in the S/N of the sample. 
We also examine the scatter in recovered fitting parameters
as a function of magnitude. Therefore, the S/N as a
function of apparent magnitude should appropriately reflect that of the parent
sample from SDSS.

\begin{figure} 
\includegraphics[width=\columnwidth]{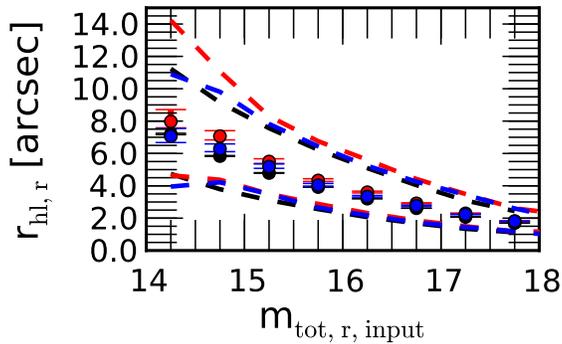}
\caption{The distribution of galaxy radii as a function of apparent magnitude
for the parent SDSS sample in black, the \Ser\ model in red and the
\SerExp\ model in blue. The median in each bin is shown with error bars
representing the 95\% CI on the median.
Corresponding dashed lines show the extent of the middle 68\% of data. The
\SerExp\ model is in close agreement across the entire magnitude range while
the \Ser\ model begins to diverge at brighter magnitudes.}
\label{fig:hrad_vs_appmag}
\end{figure}

\begin{figure} 
\includegraphics[width=\columnwidth]{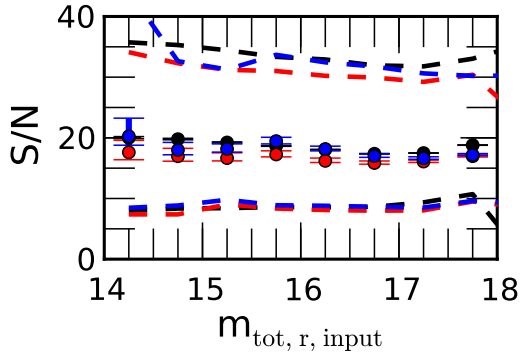}
\caption{The distribution of galaxy S/N as a function of apparent magnitude is
presented in the same format as Figure \ref{fig:hrad_vs_appmag}. The
\SerExp\ and \Ser\    
models are in close agreement with the full sample across the entire magnitude
range.}
\label{fig:SN_vs_appmag}
\end{figure}

Figure~\ref{fig:hrad_vs_appmag} presents the halflight radius versus apparent
magnitude, and Figure~\ref{fig:SN_vs_appmag}
presents the S/N versus apparent magnitude. The points
shown in red and blue correspond to the \Ser\ and \SerExp\
mocks, respectively. The underlying SDSS parent distribution is shown in black.
Figure~\ref{fig:hrad_vs_appmag} shows that the \Ser\ and \SerExp\ models
are in close agreement with the full SDSS sample. The \Ser\ and \SerExp\ model
radii agree across the magnitude range. The S/N agrees with the
full SDSS sample or is slightly below that of SDSS. The lower signal-to-noise,
although not exactly that of SDSS, will not bias the tests toward better
results, so we deem these samples acceptable for testing. The \DevExp\ sample,
which is not shown here, 
tends to have smaller radii and higher S/N at brighter magnitude. The results
of tests using the \DevExp\ model are not discussed in the remainder
of this paper. They can be found in \catalog.

\begin{figure*}
\includegraphics{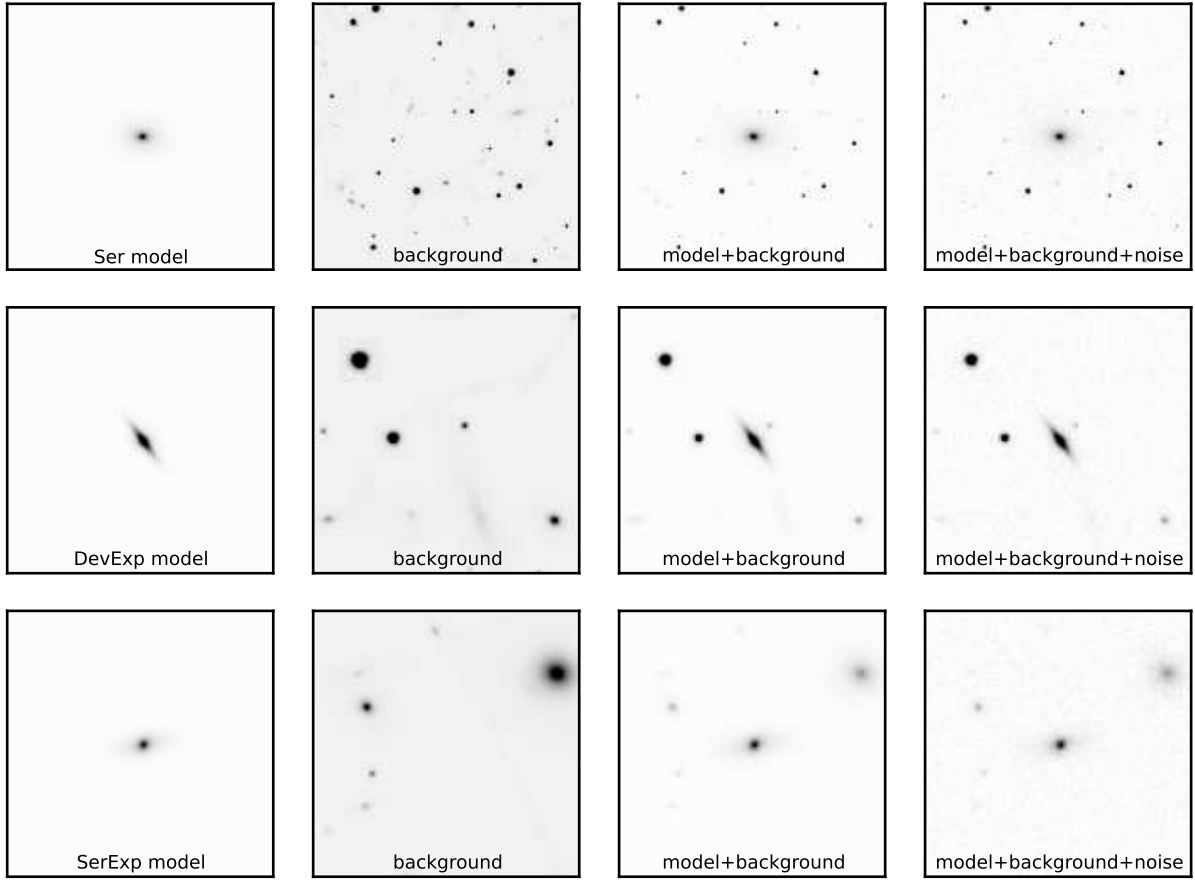}
\caption{Examples of mock galaxies and background shown before and
after adding Poisson noise. Top, middle, and bottom rows show randomly
selected sample \Ser, \DevExp, and \SerExp\ profiles, respectively. 
From left to right, the columns show the mock galaxy, simulated 
background, background+galaxy, and final image with noise.\label{fig:ex_images}} 
\end{figure*}

\subsection{Final processing for fitting}
For each mock galaxy, we also generate a weight image of the $\sigma_{i,j}$
values according to Equation~\ref{eq:weight}. This image is supplied along
with the input image to the pipeline in order to calculate the $\chi^2$
value for the fit. 

Figure \ref{fig:ex_images} shows some examples of mock galaxies 
throughout the simulation process. This includes the noiseless mock 
galaxy, the noiseless simulated background, the composite image of galaxy 
and background, and the composite image after adding Poisson noise with 
$\sigma_{i,j}$ defined in Equation~\ref{eq:weight}. The final image size 
used for fitting is 20 times the Petrosian r-band halflight radius. 
A discussion of this choice of stamp size is presented in Section~\ref{sec:cut_size}.

\section{Testing PyMorph image decompositions}\label{sec:tests}
In order to test the parameter recovery of the PyMorph
pipeline on SDSS spectroscopic galaxies, we apply the PyMorph pipeline to the
mocks described in
Section~\ref{sec:sims}. The PyMorph pipeline uses GALFIT to fit smooth profiles
to the the mock galaxies. We apply the pipeline to several different
realizations of our mock galaxies. These realizations increase in
complexity from a noiseless image to an image with real noise and
(possibly clustered) neighboring sources. We show that the ability of PyMorph
to reliably recover model parameters is limited by both the S/N and the
resolution of the mock galaxy. Understanding the systematic effects of S/N
and resolution is useful in interpreting the data presented
in \catalog. It may also be used to correct biases in the data as described
later in Section~\ref{sec:discuss}.

\subsection{Noiseless images}\label{subsec:flat}
As an initial test, the pipeline is applied to simulations prior to adding 
noise, background counts, or neighboring sources. This produces
the minimum scatter in the data, serves to verify that our simulations are
correct, and shows that PyMorph is properly functioning. 
  
\begin{figure*}

\begin{minipage}{0.23\linewidth}
\includegraphics[width=0.99\linewidth]
{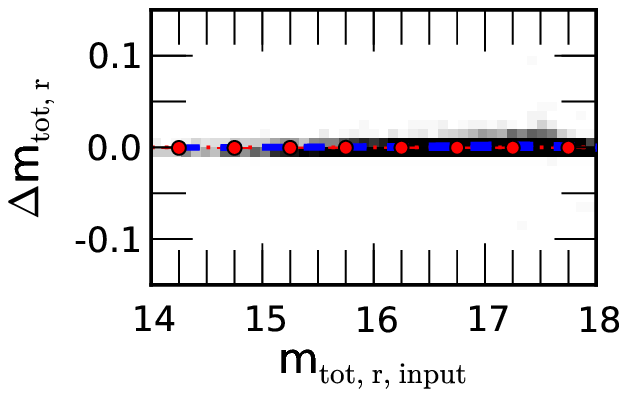}
\includegraphics[width=0.99\linewidth]
{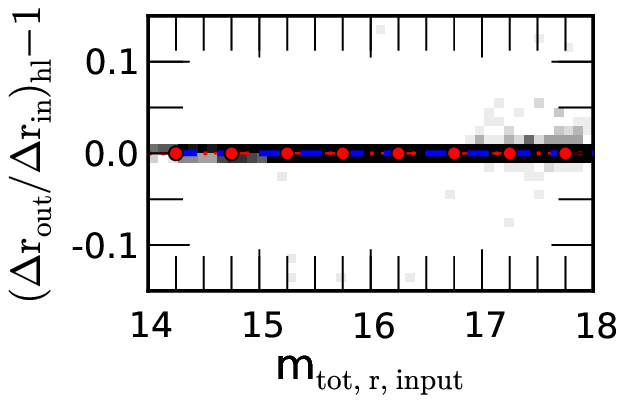}
\includegraphics[width
=0.99\linewidth]{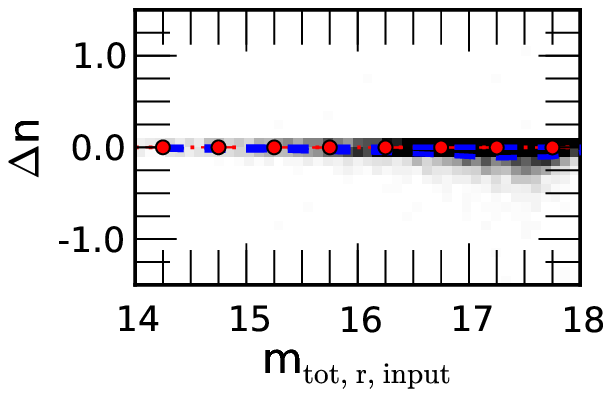}
\includegraphics[width=0.99\linewidth]
{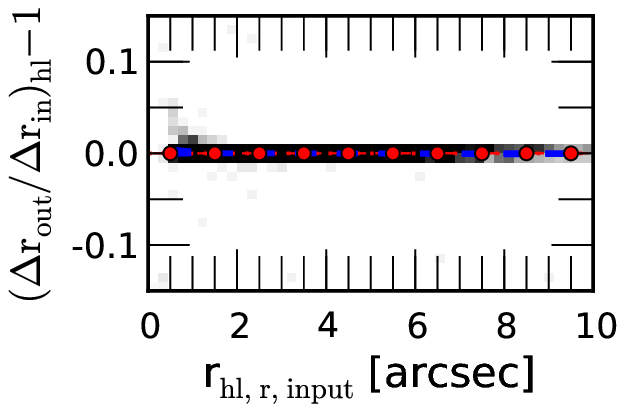}
\includegraphics[width=0.99\linewidth]
{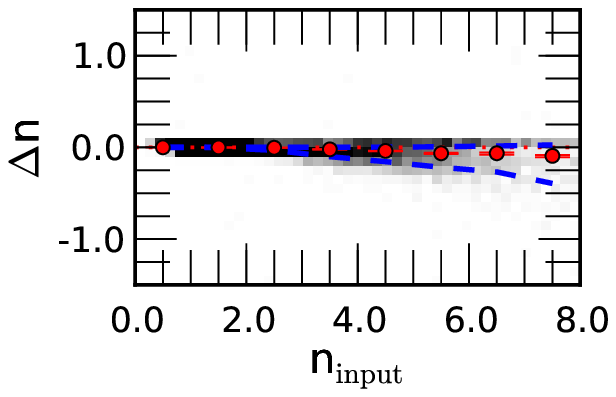}
\subcaption{}\label{subfig:flat_ser_ser}
\end{minipage}
\begin{minipage}{0.23\linewidth}
\includegraphics[width=0.99\linewidth]
{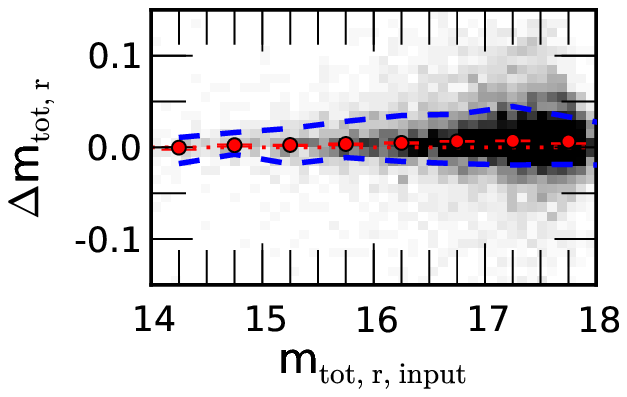}
\includegraphics[width=0.99\linewidth]
{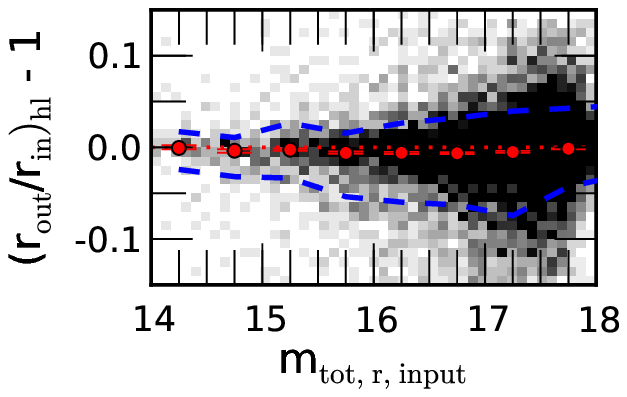}
\includegraphics[width=0.99\linewidth]
{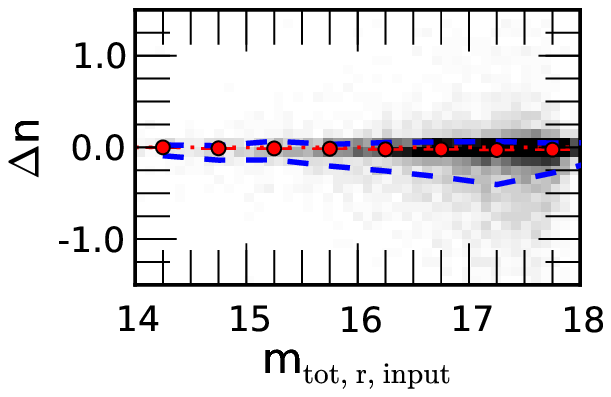}
\includegraphics[width=0.99\linewidth]
{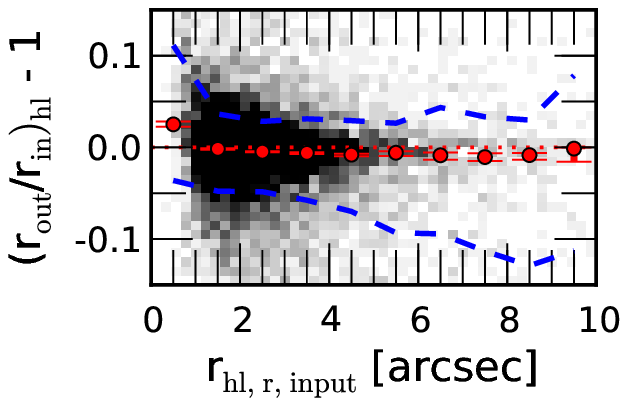}
\includegraphics[width
=0.99\linewidth]{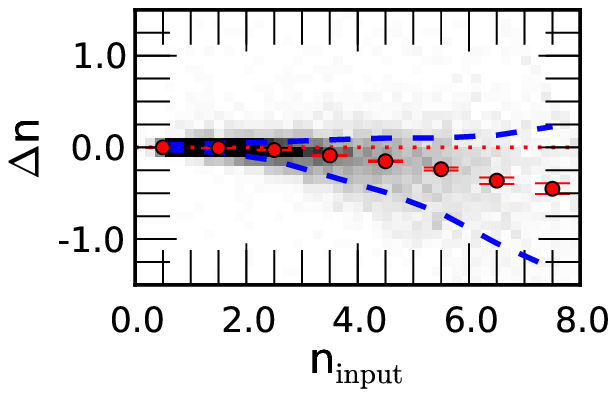}
\subcaption{}\label{subfig:psf_ser_ser}
\end{minipage}
\begin{minipage}{0.23\linewidth}
\includegraphics[width=0.99\linewidth]
{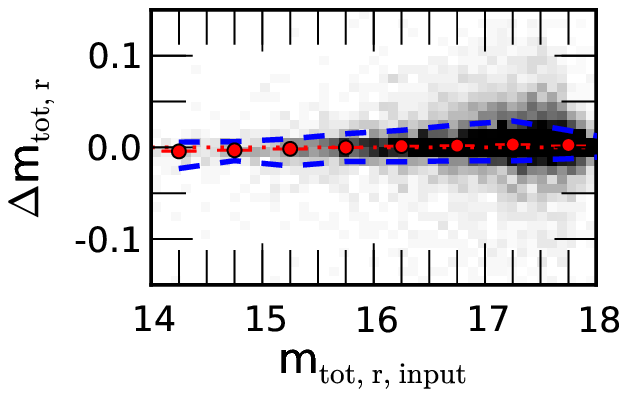}
\includegraphics[width=0.99\linewidth]
{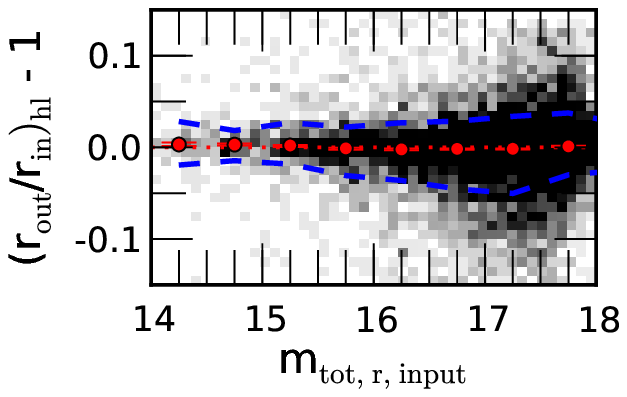}
\includegraphics[width=0.99\linewidth]
{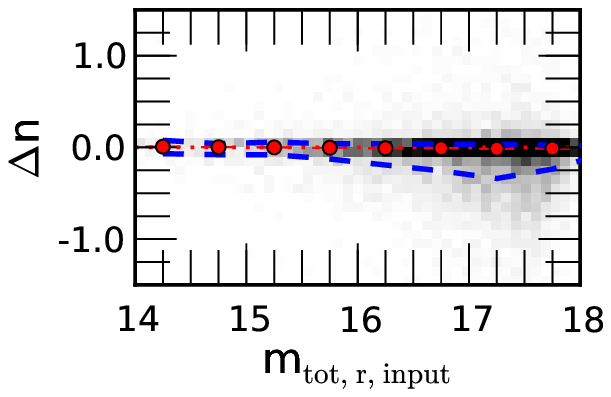}
\includegraphics[width=0.99\linewidth]
{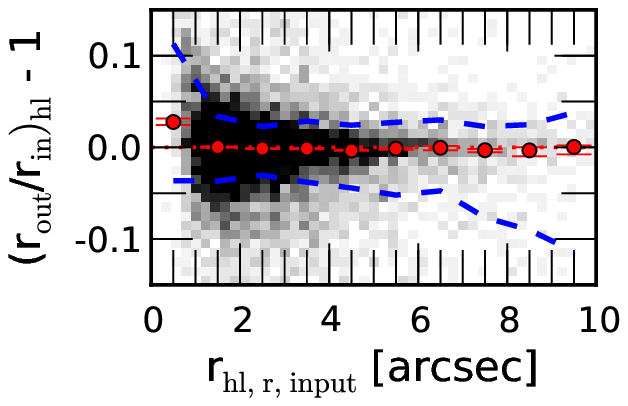}
\includegraphics[width
=0.99\linewidth]{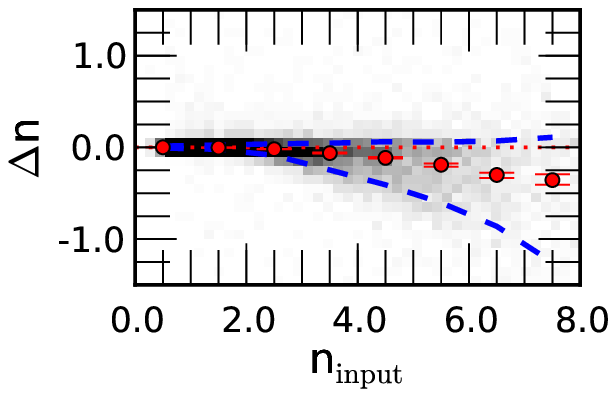}
\subcaption{}\label{subfig:sn4_ser_ser}
\end{minipage}
\begin{minipage}{0.23\linewidth}
\includegraphics[width=0.99\linewidth]
{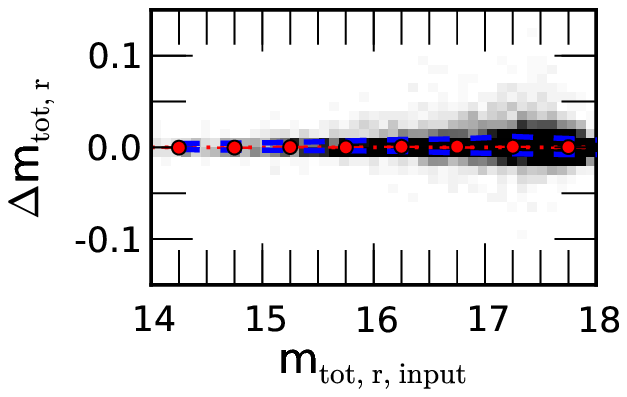}
\includegraphics[width=0.99\linewidth]
{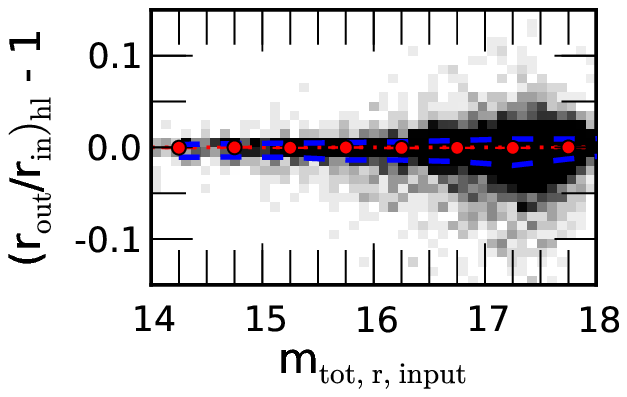}
\includegraphics[width=0.99\linewidth]
{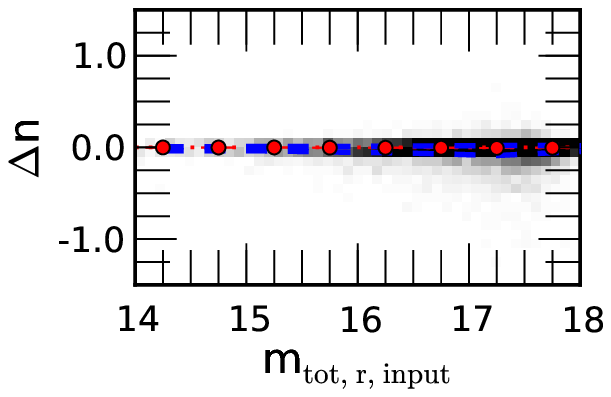}
\includegraphics[width=0.99\linewidth]
{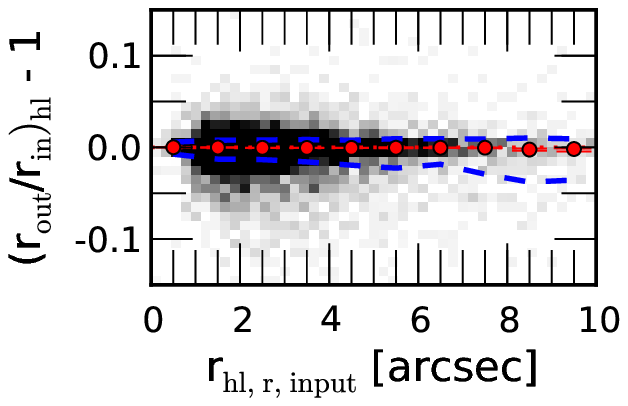}
\includegraphics[width
=0.99\linewidth]{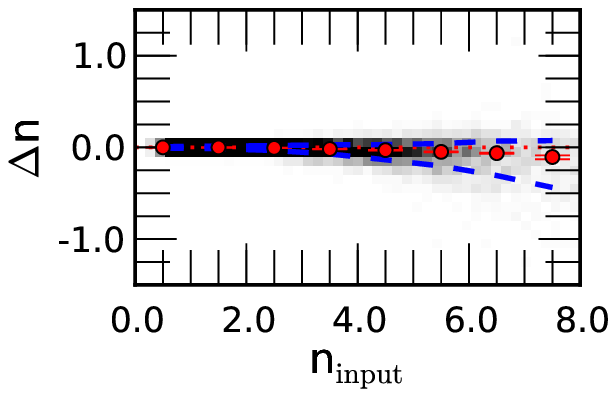}
\subcaption{}\label{subfig:pix2_ser_ser}
\end{minipage}

\caption{The simulated and recovered
apparent magnitude, halflight radius, and S{\'e}rsic index for a \Ser\ galaxy 
fit with a \Ser\ model in four cases: \textbf{\subref{subfig:flat_ser_ser}} the
image prior to adding Poisson noise, \textbf{\subref{subfig:psf_ser_ser}} our
fiducial case containing simulated sky, Poisson noise, PSF errors, and neighboring
sources,
\textbf{\subref{subfig:sn4_ser_ser}} the fiducial case with S/N increased by a
factor of 4, and \textbf{\subref{subfig:pix2_ser_ser}} the fiducial case
with resolution increased by a factor of 2. 
Over-plotted are the bias (red points) in the fitted values. 
All plots show the 68\% (dashed line) scatter in blue. The density
of points is plotted in gray-scale. The S{\'e}rsic index shows increasing
underestimate up to $\approx$0.5 (or $\approx$6\%) at the largest S{\'e}rsic
indexes.}
\label{fig:ser_ser}
\end{figure*}

\begin{figure*}

\begin{minipage}{0.23\linewidth}
\includegraphics[width=0.99\linewidth]
{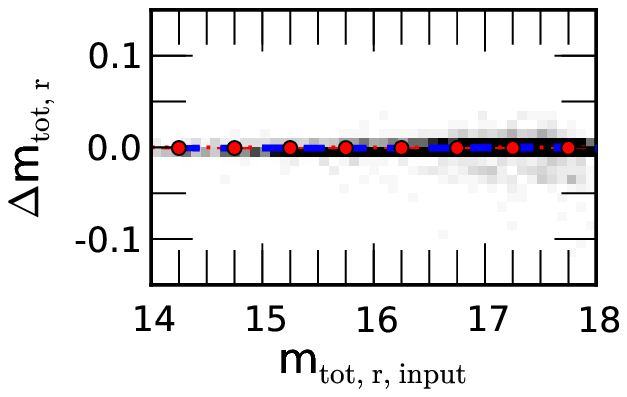}
\includegraphics[width=0.99\linewidth]
{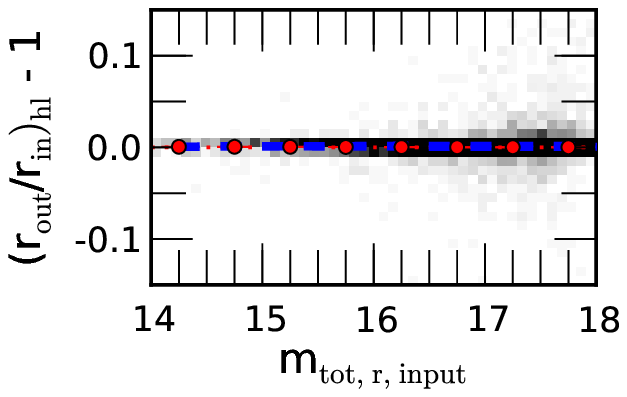}
\includegraphics[width=0.99\linewidth]
{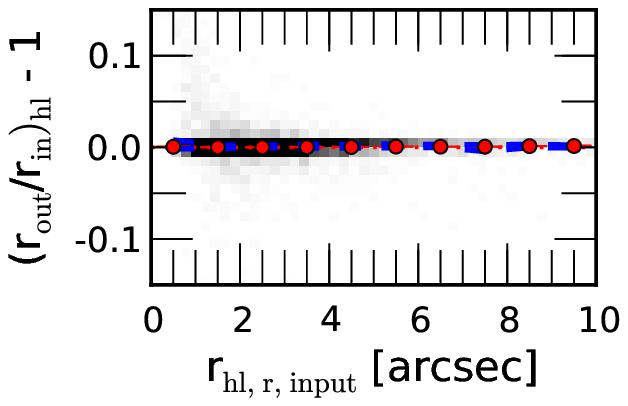}
\subcaption{}\label{subfig:flat_ser_serexp}
\end{minipage}
\begin{minipage}{0.23\linewidth}
\includegraphics[width=0.99\linewidth]
{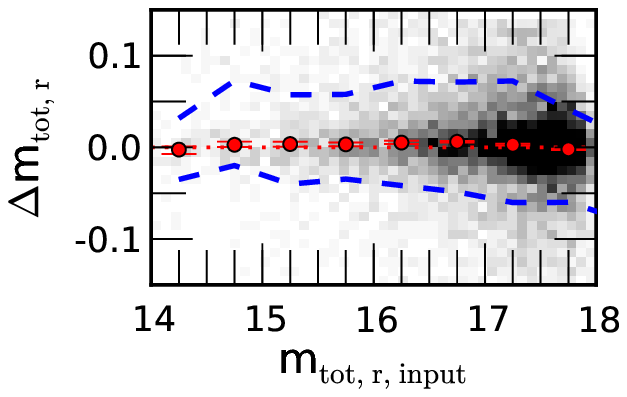}
\includegraphics[width=0.99\linewidth]
{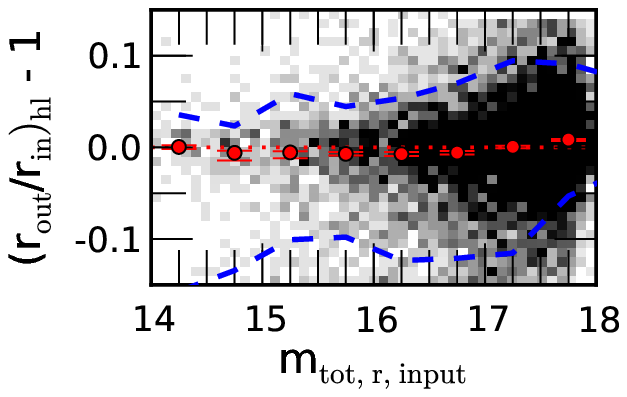}
\includegraphics[width=0.99\linewidth]
{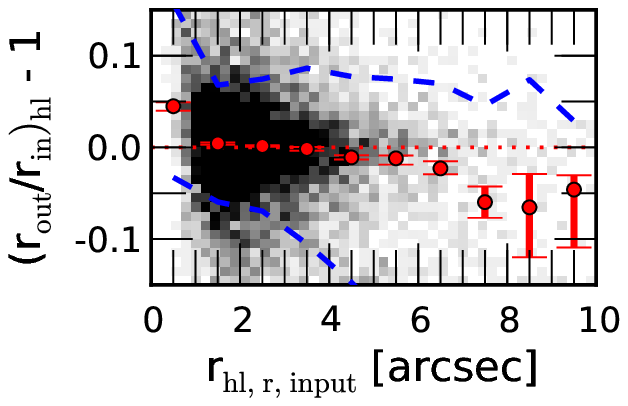}
\subcaption{}\label{subfig:psf_ser_serexp}
\end{minipage}
\begin{minipage}{0.23\linewidth}
\includegraphics[width=0.99\linewidth]
{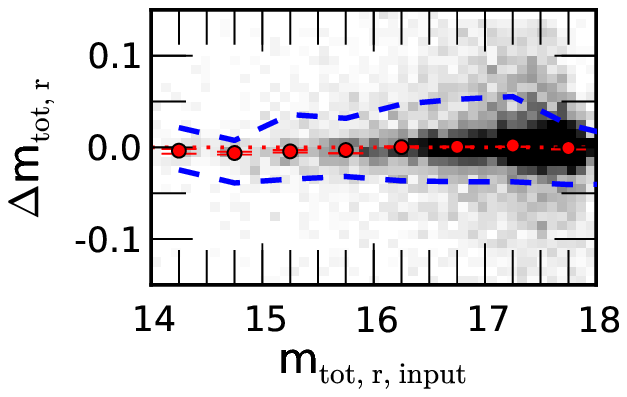}
\includegraphics[width=0.99\linewidth]
{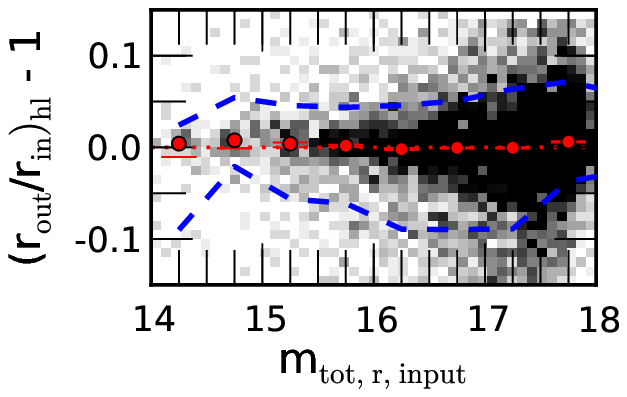}
\includegraphics[width=0.99\linewidth]
{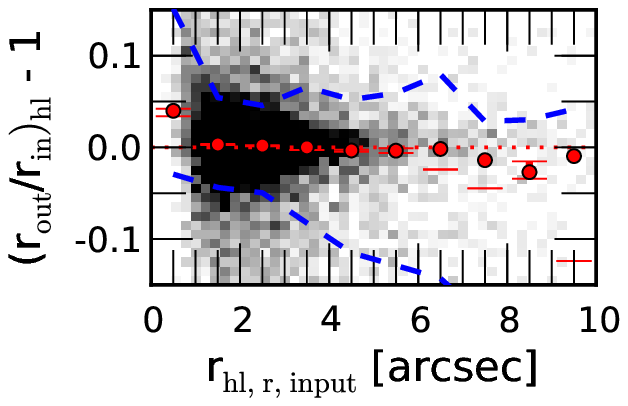}
\subcaption{}\label{subfig:sn4_ser_serexp}
\end{minipage}
\begin{minipage}{0.23\linewidth}
\includegraphics[width=0.99\linewidth]
{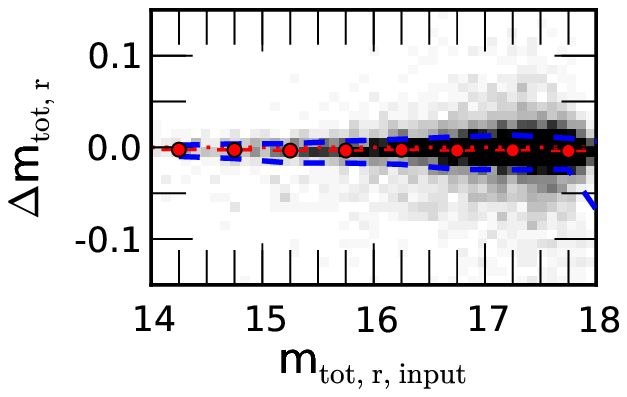}
\includegraphics[width=0.99\linewidth]
{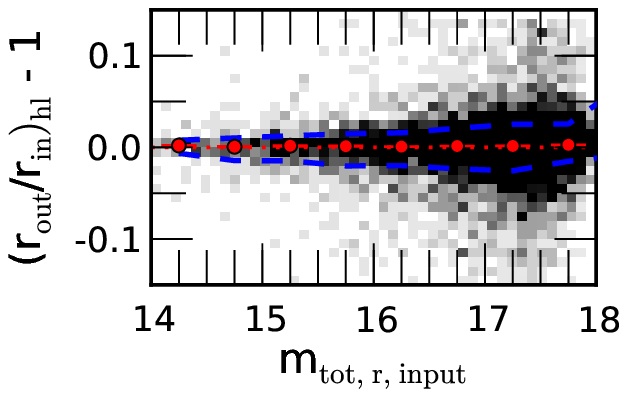}
\includegraphics[width=0.99\linewidth]
{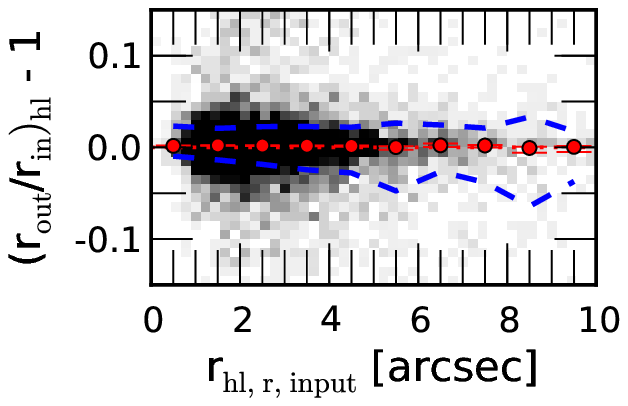}
\subcaption{}\label{subfig:pix2_ser_serexp}
\end{minipage}

\caption{The simulated and recovered
apparent magnitude and halflight radius for a \Ser\ galaxy 
fit with a \SerExp\ model in four cases: 
\textbf{\subref{subfig:flat_ser_serexp}} the
image prior to adding Poisson noise, \textbf{\subref{subfig:psf_ser_serexp}} our
fiducial case containing simulated sky, Poisson noise, PSF errors, and neighboring
sources,
\textbf{\subref{subfig:sn4_ser_serexp}} the fiducial case with S/N increased by
a
factor of 4, and \textbf{\subref{subfig:pix2_ser_serexp}} the fiducial case
with resolution increased by a factor of 2.  
Over-plotted are the bias (red points) in the fitted values. 
 All plots show the 68\% (dashed line) scatter in blue. The density
of points is plotted in gray-scale.
}
\label{fig:ser_serexp}

\end{figure*}

\begin{figure*}

\begin{minipage}{0.23\linewidth}
\includegraphics[width=0.99\linewidth]
{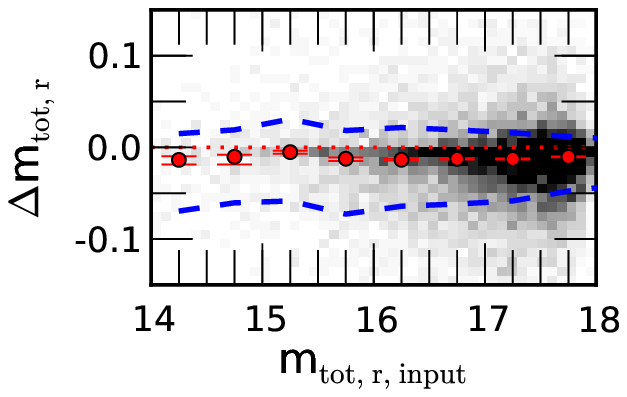}
\includegraphics[width=0.99\linewidth]
{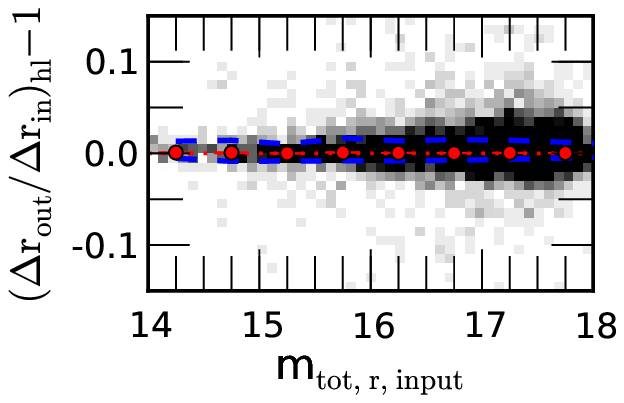}
\includegraphics[width=0.99\linewidth]
{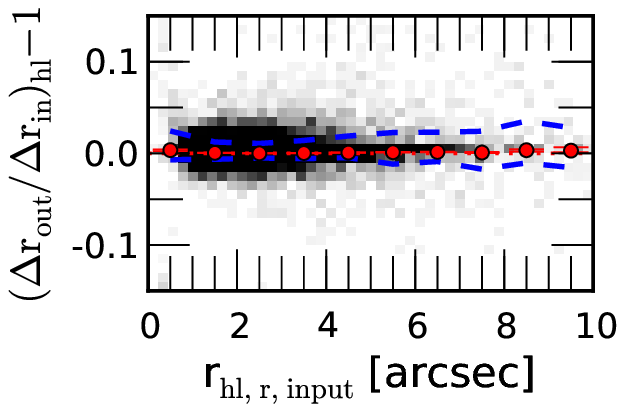}
\subcaption{}\label{subfig:flat_serexp_ser}
\end{minipage}
\begin{minipage}{0.23\linewidth}
\includegraphics[width=0.99\linewidth]
{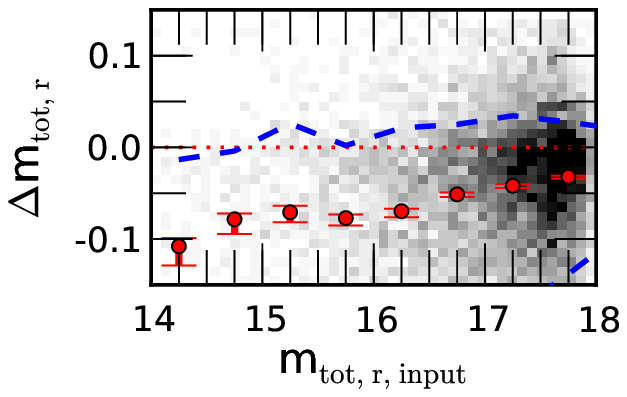}
\includegraphics[width=0.99\linewidth]
{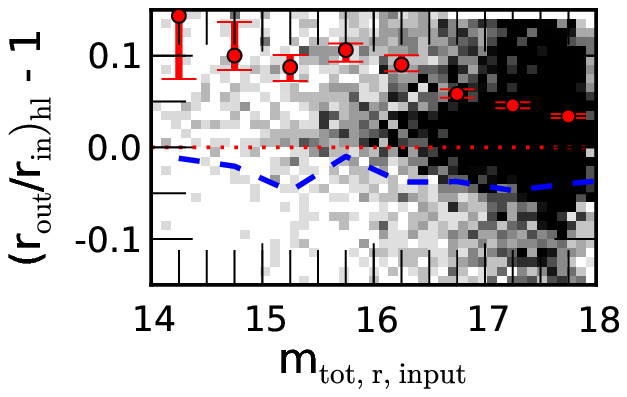}
\includegraphics[width=0.99\linewidth]
{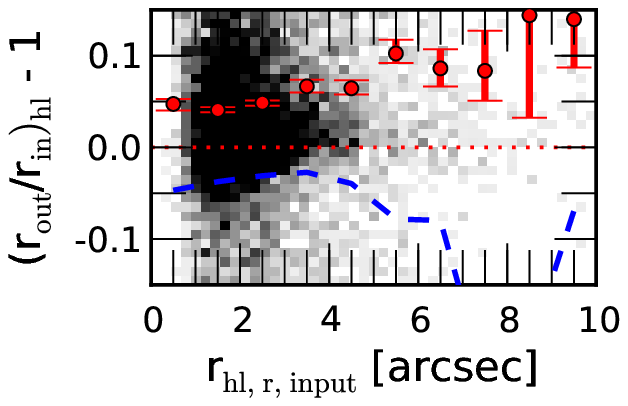}
\subcaption{}\label{subfig:psf_serexp_ser}
\end{minipage}
\begin{minipage}{0.23\linewidth}
\includegraphics[width=0.99\linewidth]
{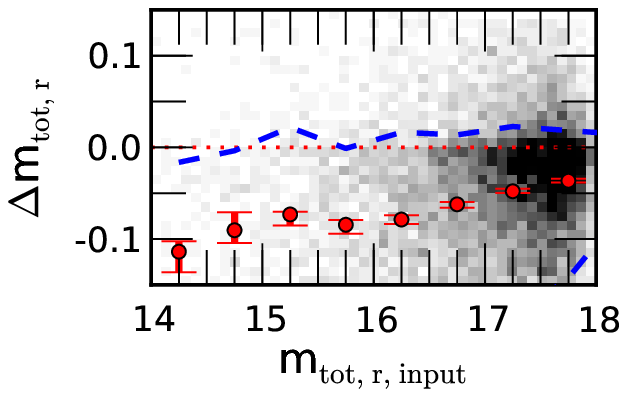}
\includegraphics[width=0.99\linewidth]
{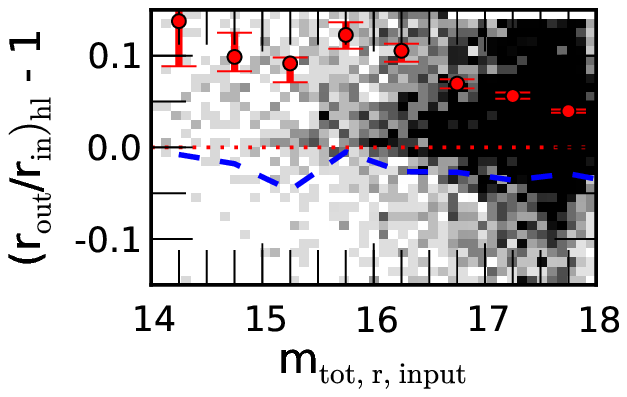}
\includegraphics[width=0.99\linewidth]
{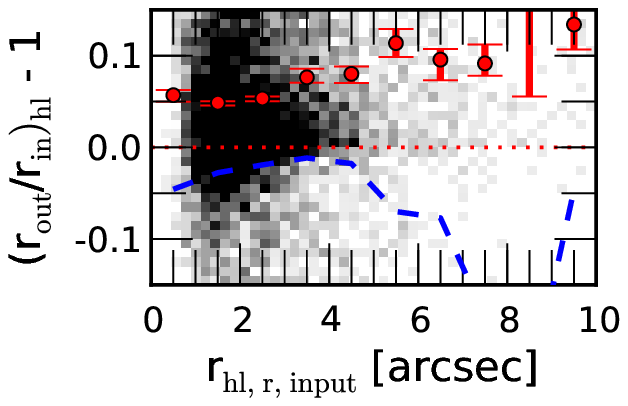}
\subcaption{}\label{subfig:sn4_serexp_ser}
\end{minipage}
\begin{minipage}{0.23\linewidth}
\includegraphics[width=0.99\linewidth]
{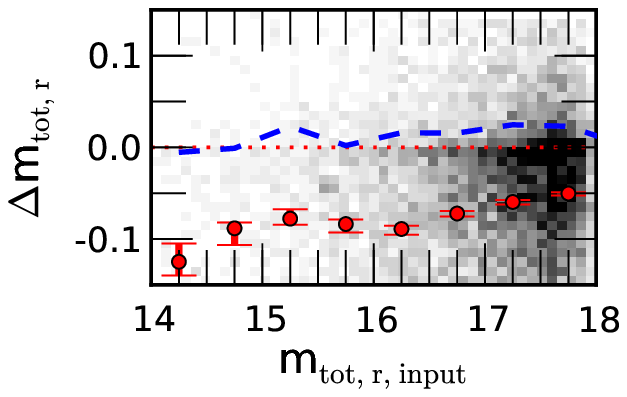}
\includegraphics[width=0.99\linewidth]
{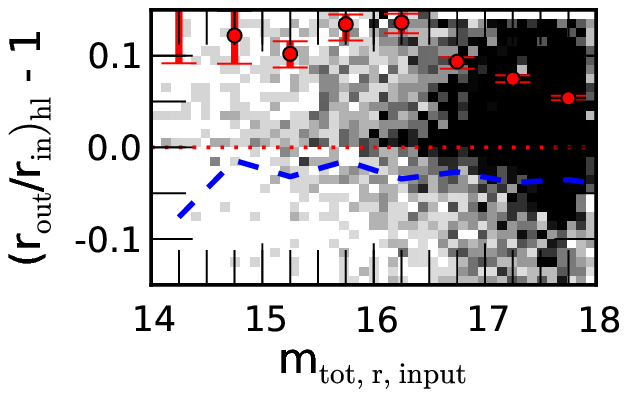}
\includegraphics[width=0.99\linewidth]
{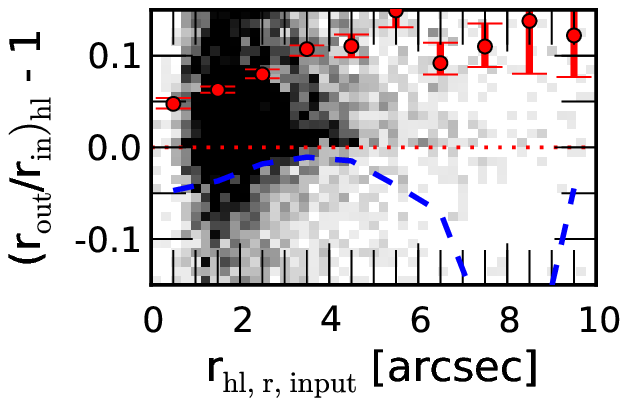}
\subcaption{}\label{subfig:pix2_serexp_ser}
\end{minipage}

\caption{The simulated and recovered
apparent magnitude and halflight radius for a \SerExp\ galaxy 
fit with a \Ser\ model in four cases:  \textbf{\subref{subfig:flat_serexp_ser}}
the image prior to adding Poisson noise, \textbf{\subref{subfig:psf_serexp_ser}}
our fiducial case containing simulated sky, Poisson noise, PSF errors, and neighboring
sources,
\textbf{\subref{subfig:sn4_serexp_ser}} the fiducial case with S/N increased by
a factor of 4, and \textbf{\subref{subfig:pix2_serexp_ser}} the fiducial case
with resolution increased by a factor of 2.  
Over-plotted are the bias (red points) in the fitted values. 
All plots show the 68\% (dashed line) scatter in blue. The density
of points is plotted in gray-scale. The inability of the \Ser\ profile to
accurately model a \SerExp\ galaxy is clearly evident. Errors in magnitude and
halflight radius are correlated and the error in radius is largely driven by
errors in the largest, brightest objects. However, systematic errors occur even
at the dimmer magnitudes. \Ser\ fits tend toward recovering larger, brighter
objects when applied to a true two component galaxy. 
}
\label{fig:serexp_ser}

\end{figure*}

\begin{figure*}

\begin{minipage}{0.23\linewidth}
\includegraphics[width=0.99\linewidth]
{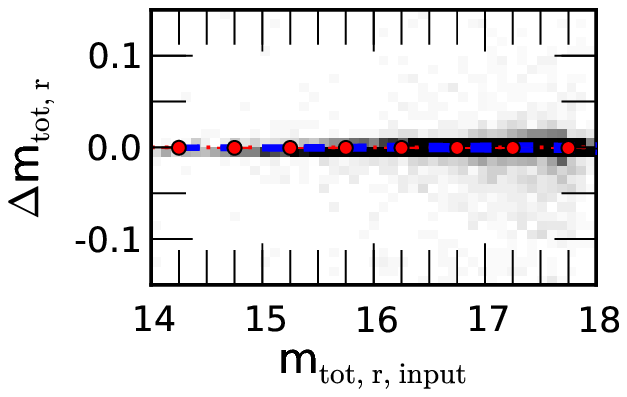}
\includegraphics[width=0.99\linewidth]
{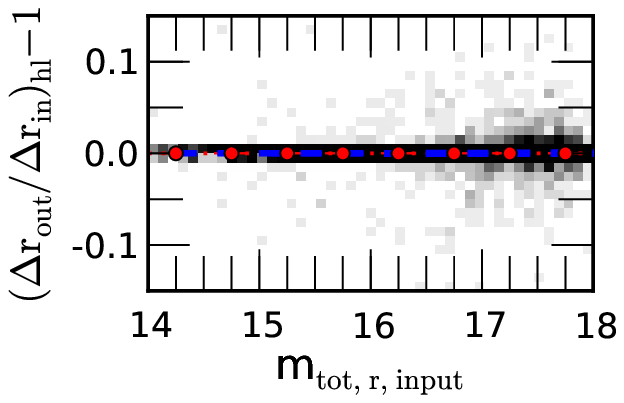}
\includegraphics[width=0.99\linewidth]
{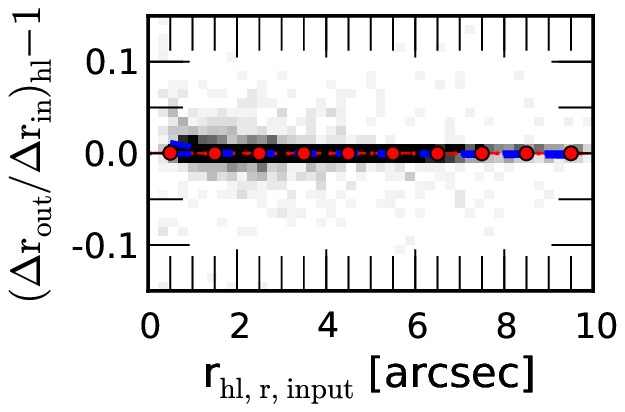}
\includegraphics[width=0.99\linewidth]
{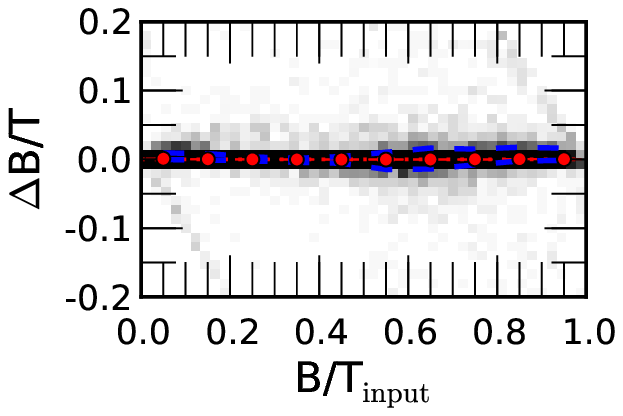}
\subcaption{}\label{subfig:flat_serexp_serexp}
\end{minipage}
\begin{minipage}{0.23\linewidth}
\includegraphics[width=0.99\linewidth]
{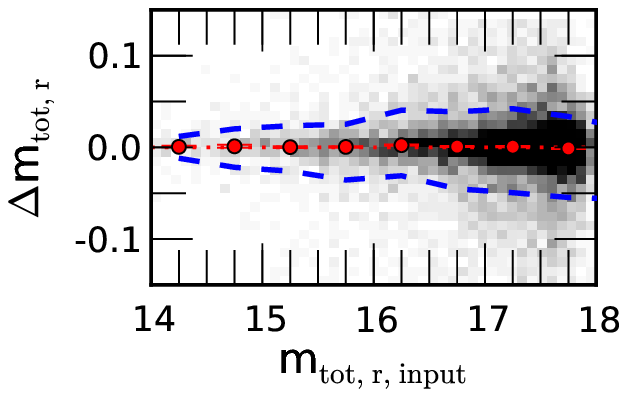}
\includegraphics[width=0.99\linewidth]
{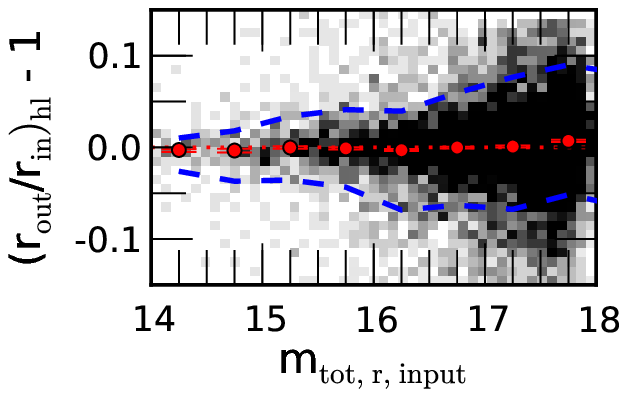}
\includegraphics[width=0.99\linewidth]
{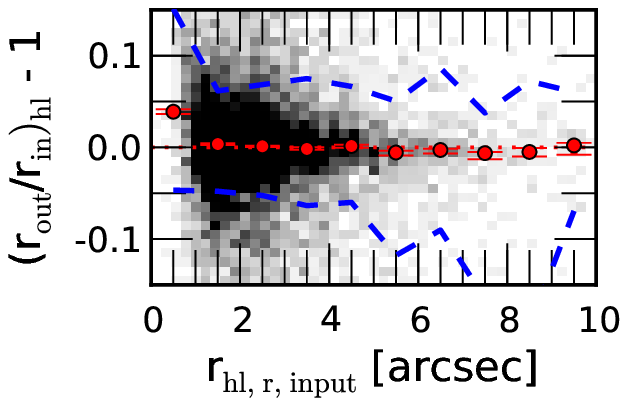}
\includegraphics[width=0.99\linewidth]
{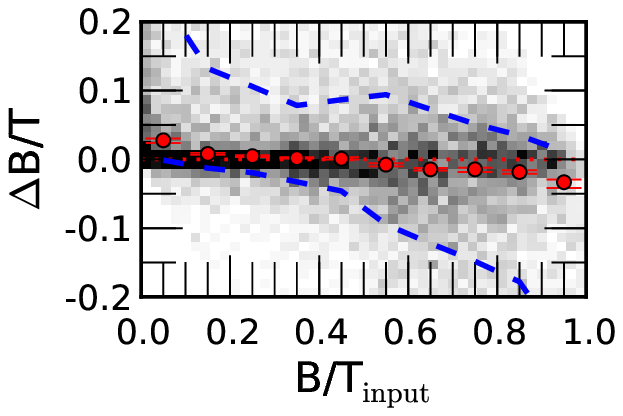}
\subcaption{}\label{subfig:psf_serexp_serexp}
\end{minipage}
\begin{minipage}{0.23\linewidth}
\includegraphics[width=0.99\linewidth]
{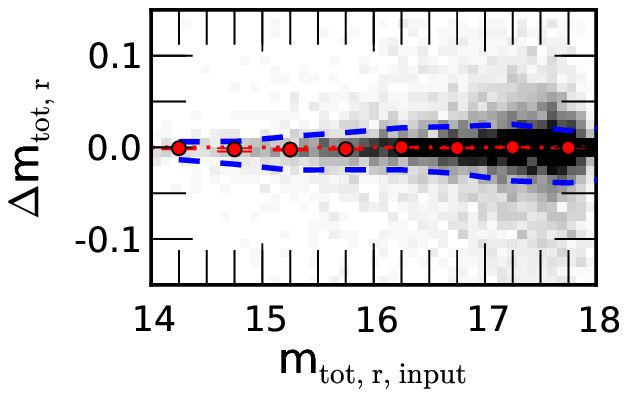}
\includegraphics[width=0.99\linewidth]
{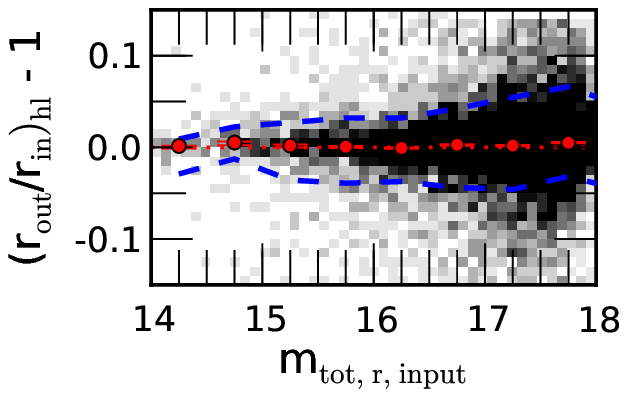}
\includegraphics[width=0.99\linewidth]
{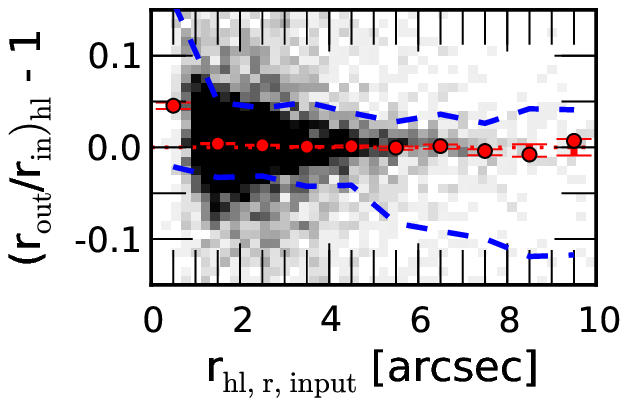}
\includegraphics[width=0.99\linewidth]
{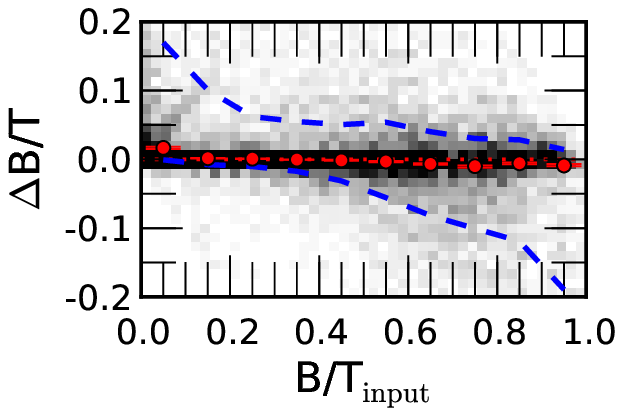}
\subcaption{}\label{subfig:sn4_serexp_serexp}
\end{minipage}
\begin{minipage}{0.23\linewidth}
\includegraphics[width=0.99\linewidth]
{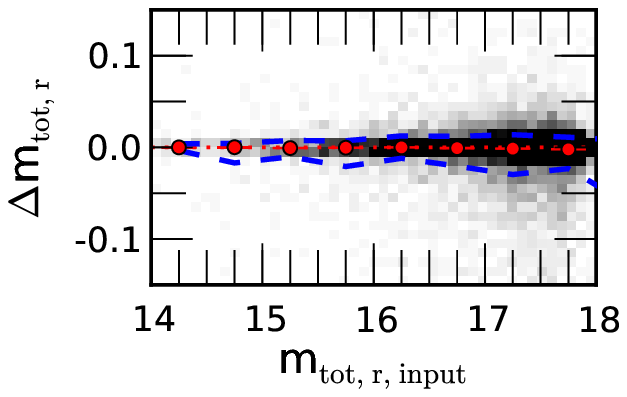}
\includegraphics[width=0.99\linewidth]
{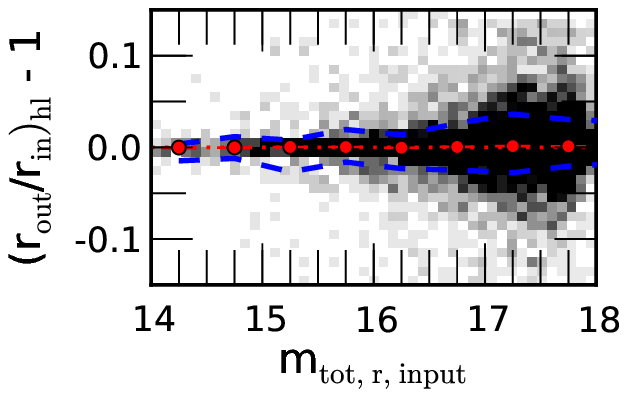}
\includegraphics[width=0.99\linewidth]
{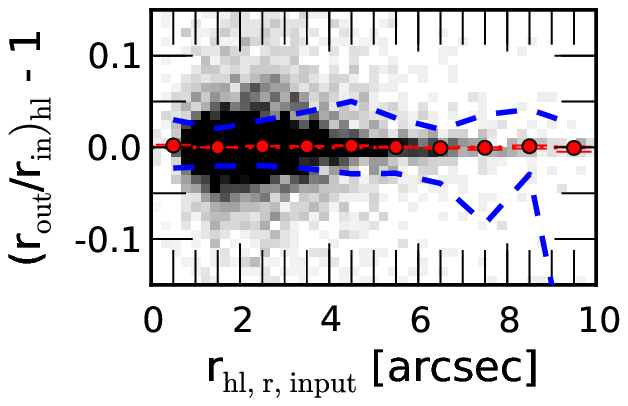}
\includegraphics[width=0.99\linewidth]
{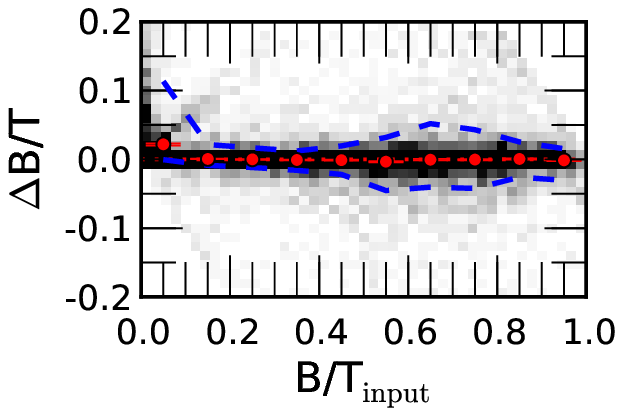}
\subcaption{}\label{subfig:pix2_serexp_serexp}
\end{minipage}

\caption{The simulated and recovered
apparent magnitude, halflight radius, and bulge-to-total light ratio for a
\SerExp\ galaxy 
fit with a \SerExp\ model in four cases:
\textbf{\subref{subfig:flat_serexp_serexp}}
the
image prior to adding Poisson noise, \textbf{\subref{subfig:psf_serexp_serexp}}
our fiducial case
containing simulated sky, Poisson noise, PSF errors, and neighboring sources,
\textbf{\subref{subfig:sn4_serexp_serexp}} the fiducial case with S/N increased
by
a factor of 4, and \textbf{\subref{subfig:pix2_serexp_ser}} the fiducial case
with resolution increased by a factor of 2. 
Over-plotted are the bias (red points) in the fitted values. 
All plots show the 68\% (dashed line) scatter in blue. The density
of points is plotted in gray-scale. The apparent trend in B/T from
overestimation at lower B/T values to underestimation at higher B/T values is
largely due to the boundaries on the parameter space forcing the scatter to
be asymmetric (\eg a galaxy with true B/T$=1$ cannot be estimated to have
B/T$>1$).
}
\label{fig:serexp_serexp}

\end{figure*}

\begin{figure*}

\begin{minipage}{0.23\linewidth}
\includegraphics[width=0.99\linewidth]
{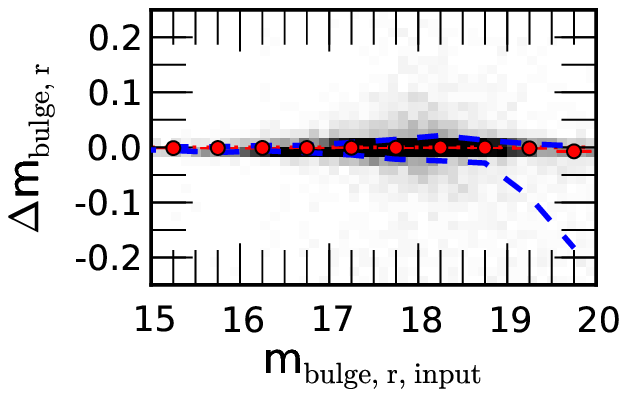}
\includegraphics[width=0.99\linewidth]
{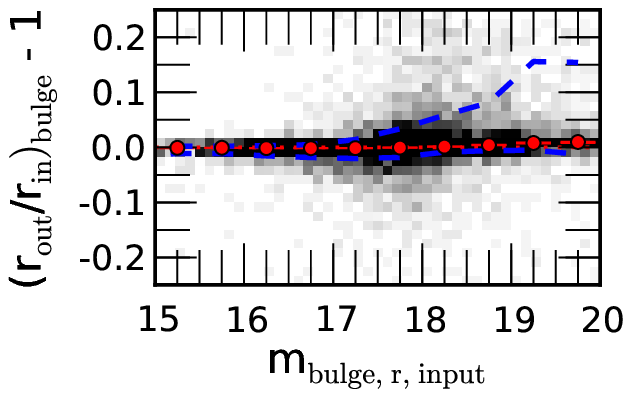}
\includegraphics[width=0.99\linewidth]
{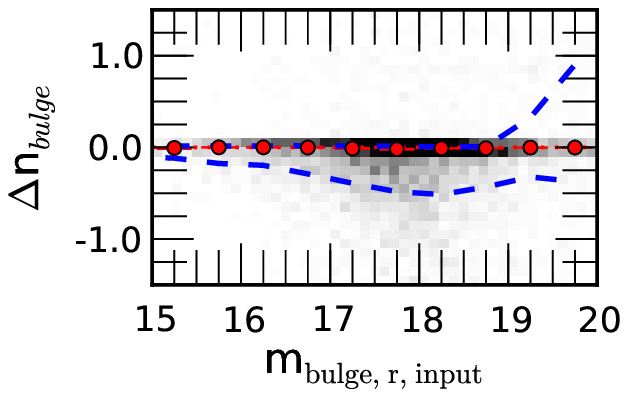}
\includegraphics[width=0.99\linewidth]
{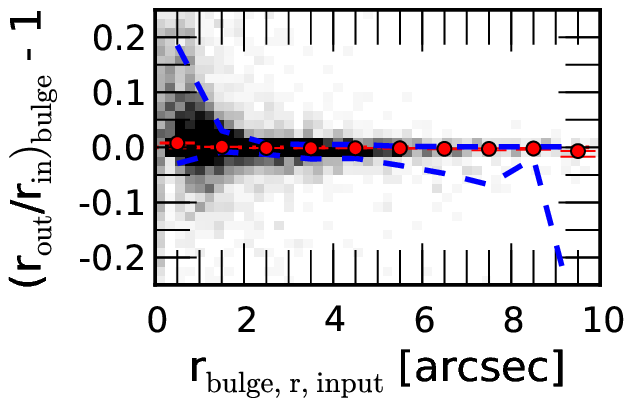}
\includegraphics[width=0.99\linewidth]
{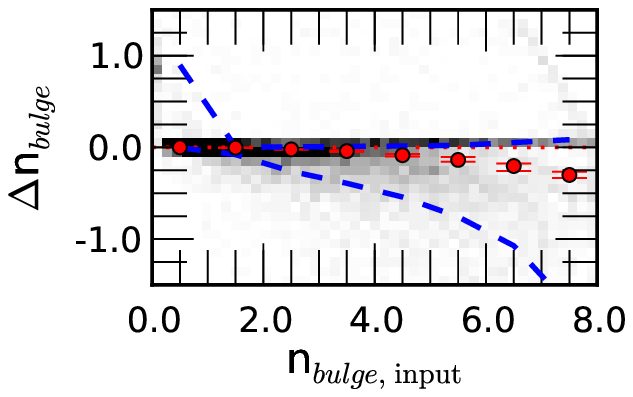}
\includegraphics[width=0.99\linewidth]
{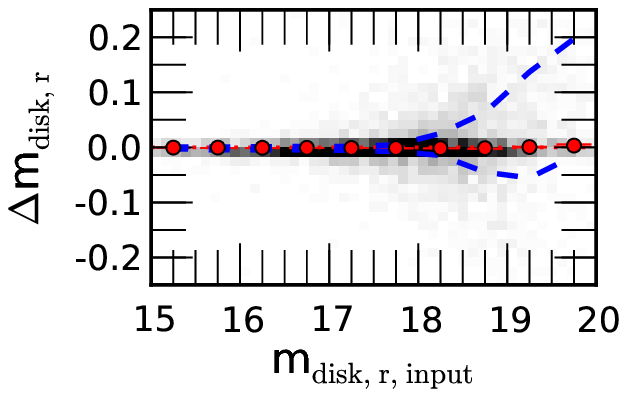}
\includegraphics[width=0.99\linewidth]
{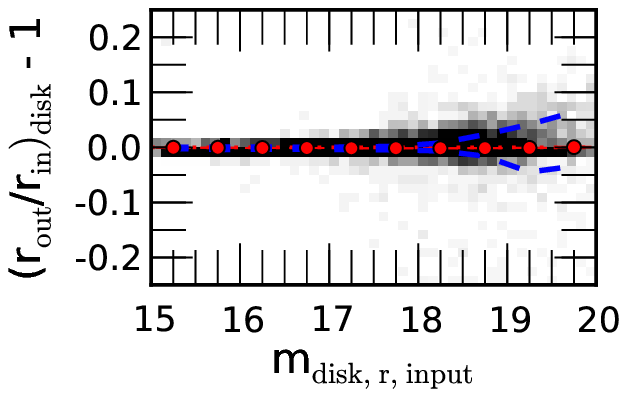}
\includegraphics[width=0.99\linewidth]
{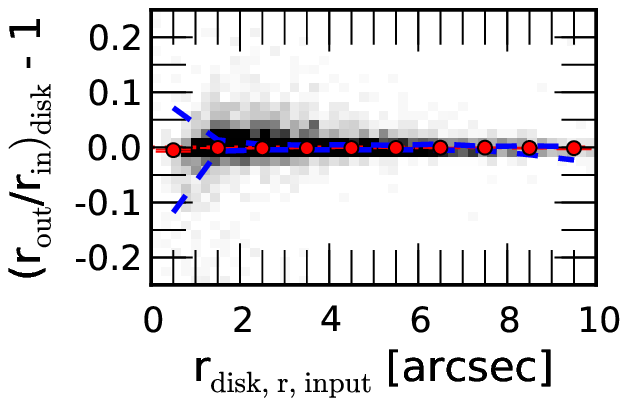}
\subcaption{}\label{subfig:flat_serexp_serexp_2com}
\end{minipage}
\begin{minipage}{0.23\linewidth}
\includegraphics[width=0.99\linewidth]
{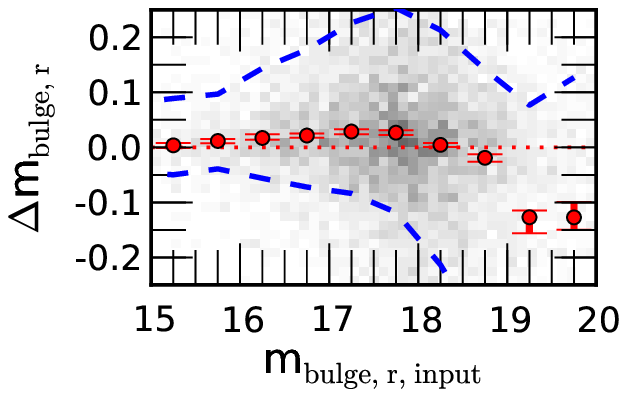}
\includegraphics[width=0.99\linewidth]
{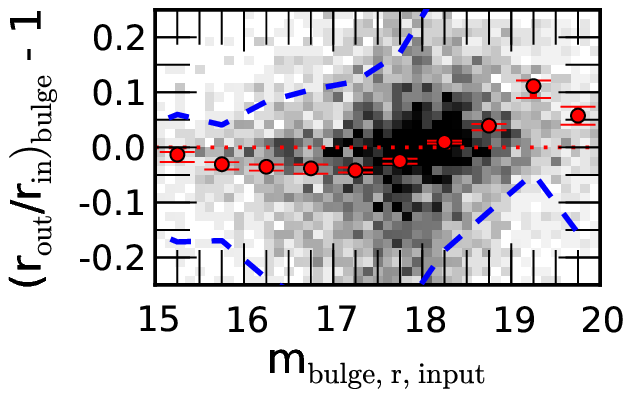}
\includegraphics[width=0.99\linewidth]
{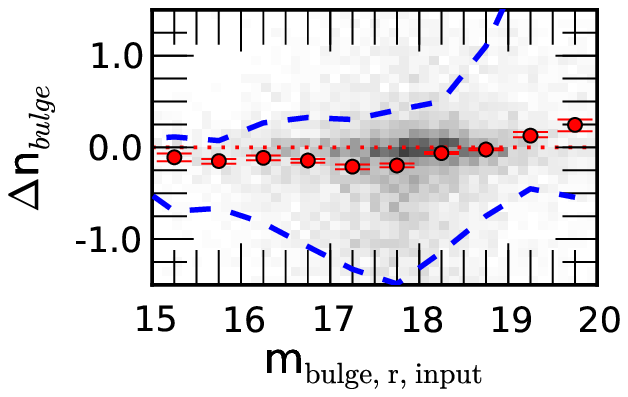}
\includegraphics[width=0.99\linewidth]
{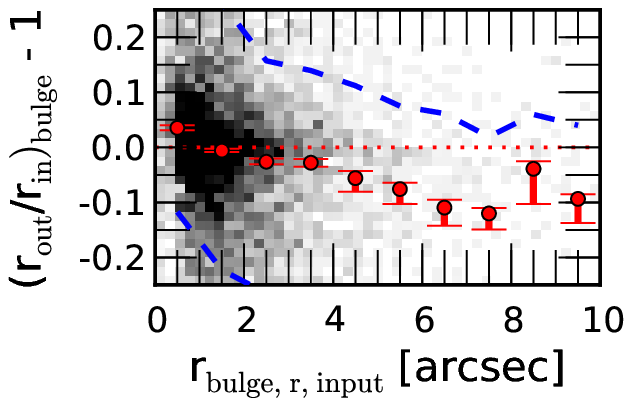}
\includegraphics[width=0.99\linewidth]
{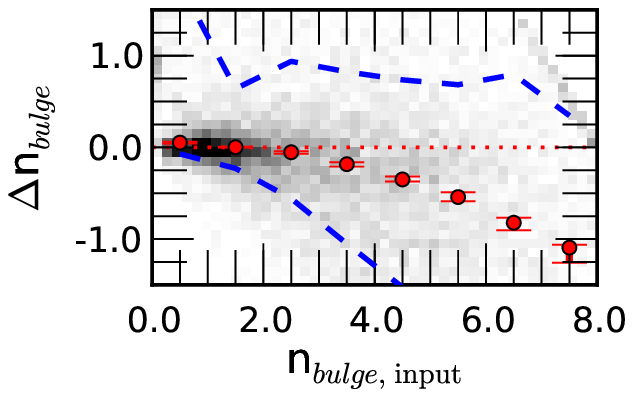}
\includegraphics[width=0.99\linewidth]
{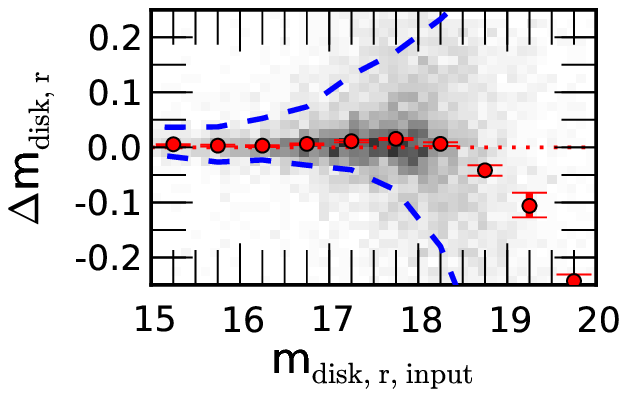}
\includegraphics[width=0.99\linewidth]
{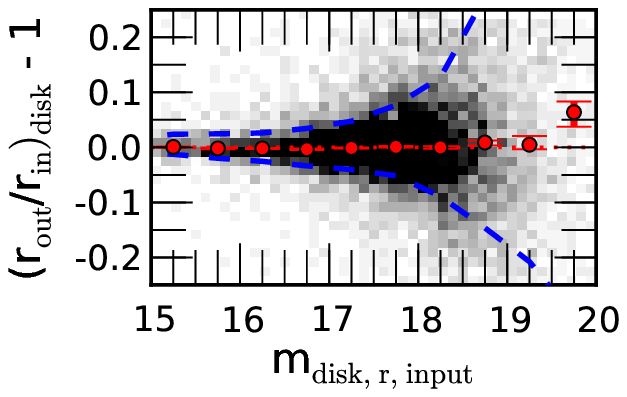}
\includegraphics[width=0.99\linewidth]
{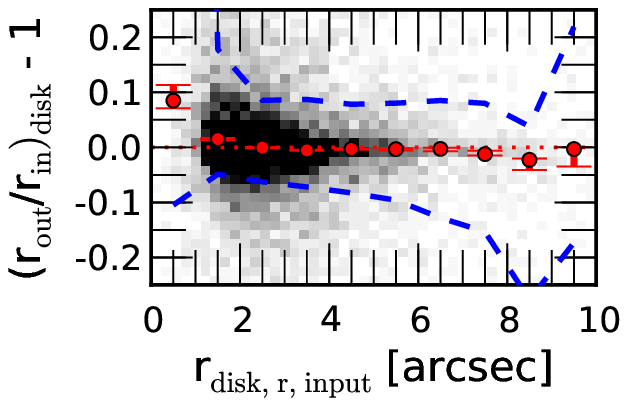}
\subcaption{}\label{subfig:psf_serexp_serexp_2com}
\end{minipage}
\begin{minipage}{0.23\linewidth}
\includegraphics[width=0.99\linewidth]
{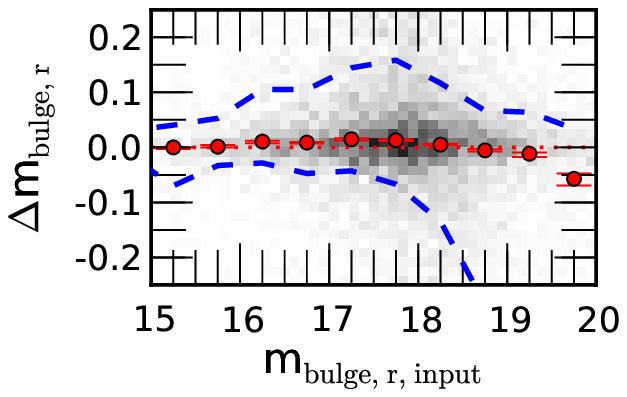}
\includegraphics[width=0.99\linewidth]
{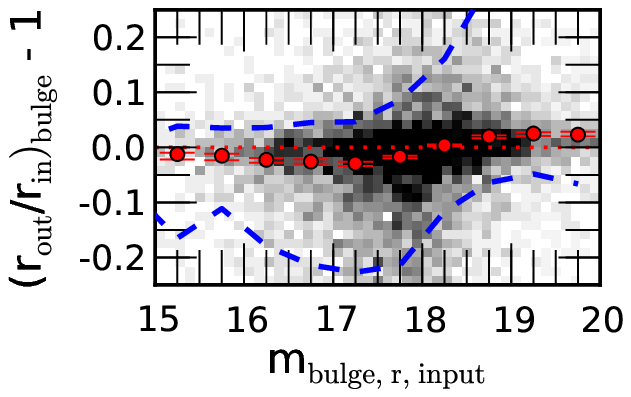}
\includegraphics[width=0.99\linewidth]
{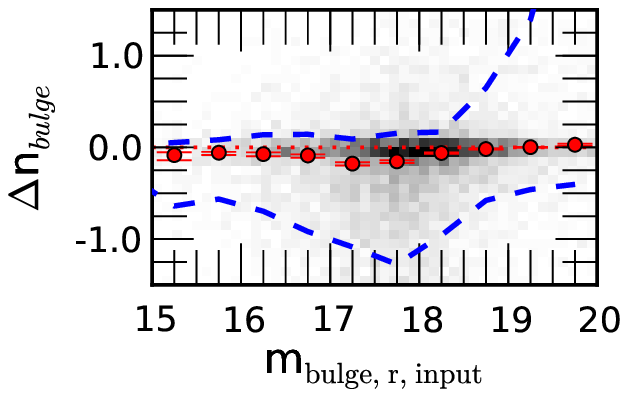}
\includegraphics[width=0.99\linewidth]
{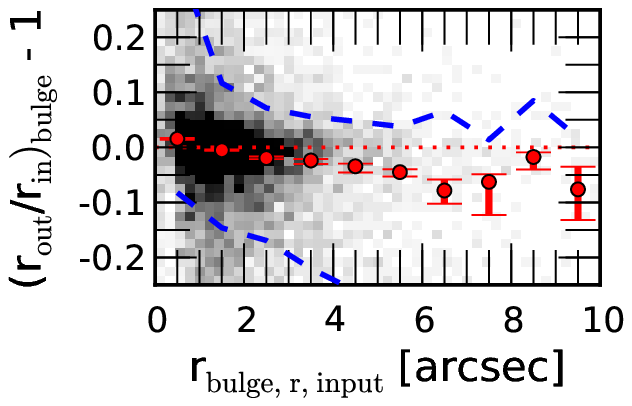}
\includegraphics[width=0.99\linewidth]
{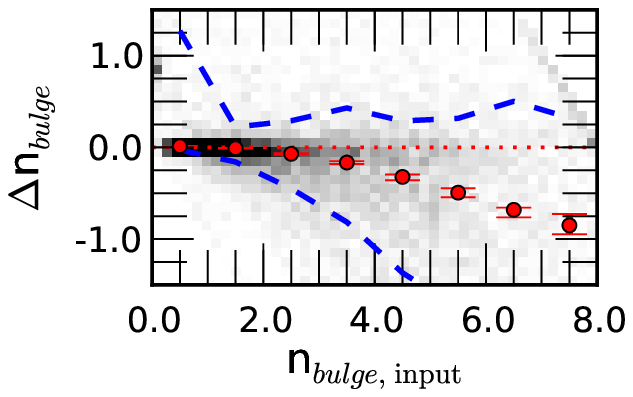}
\includegraphics[width=0.99\linewidth]
{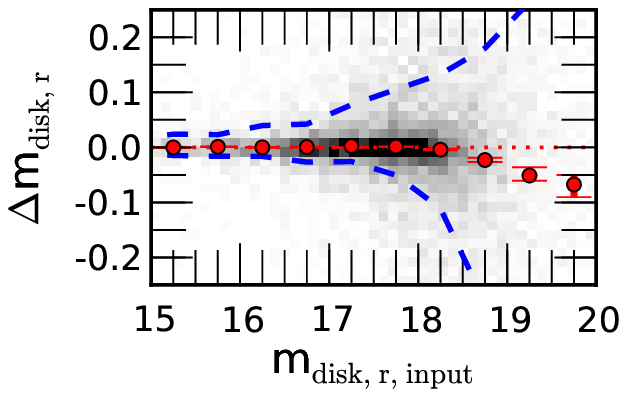}
\includegraphics[width=0.99\linewidth]
{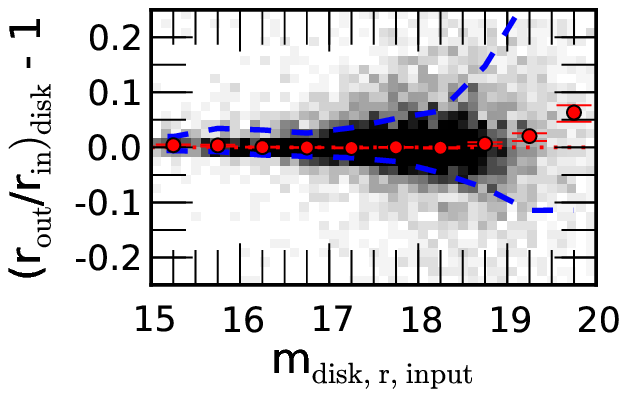}
\includegraphics[width=0.99\linewidth]
{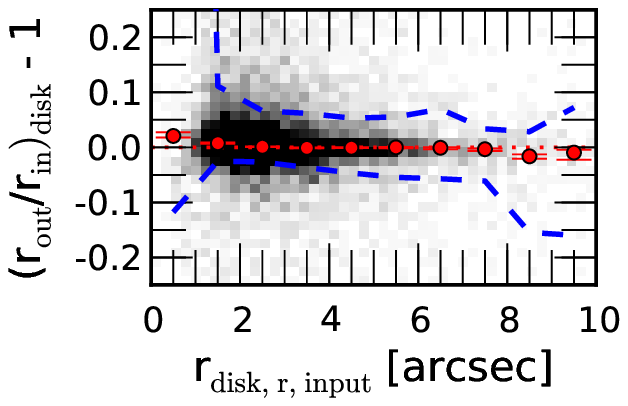}
\subcaption{}\label{subfig:sn4_serexp_serexp_2com}
\end{minipage}
\begin{minipage}{0.23\linewidth}
\includegraphics[width=0.99\linewidth]
{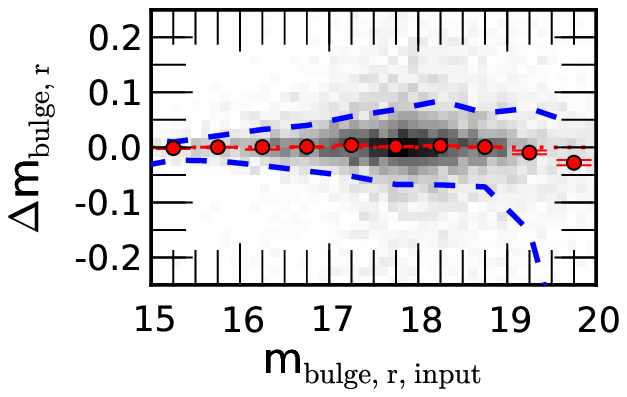}
\includegraphics[width=0.99\linewidth]
{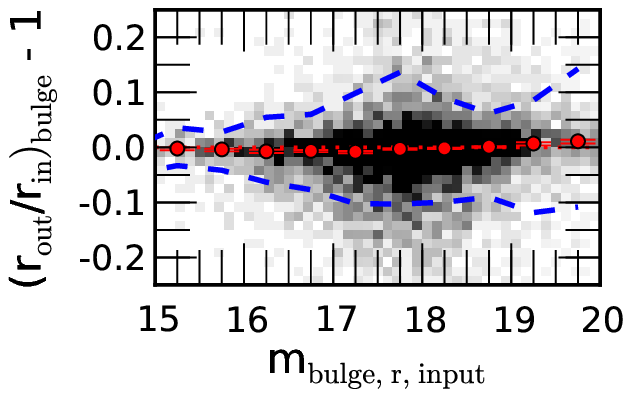}
\includegraphics[width=0.99\linewidth]
{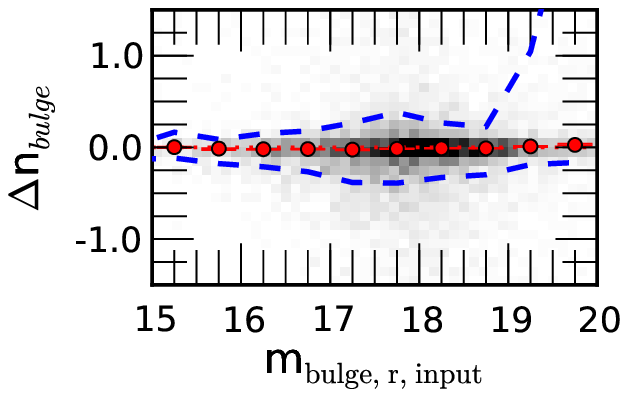}
\includegraphics[width=0.99\linewidth]
{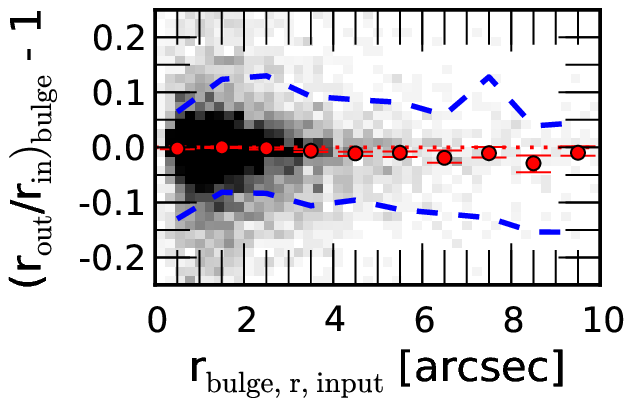}
\includegraphics[width=0.99\linewidth]
{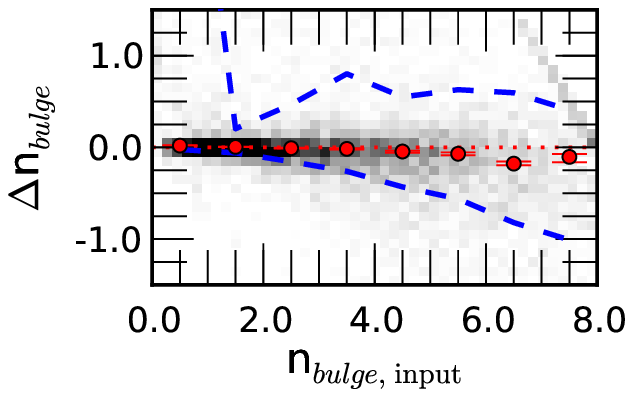}
\includegraphics[width=0.99\linewidth]
{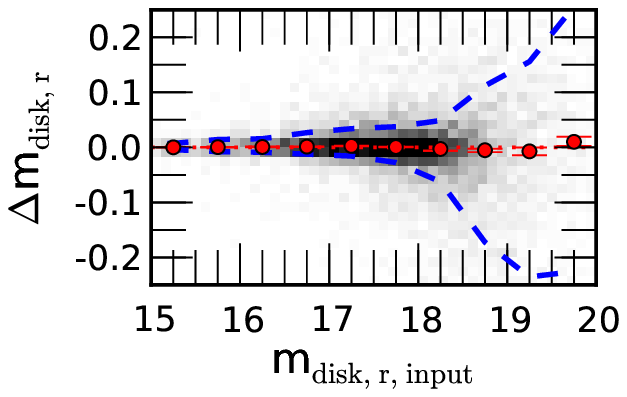}
\includegraphics[width=0.99\linewidth]
{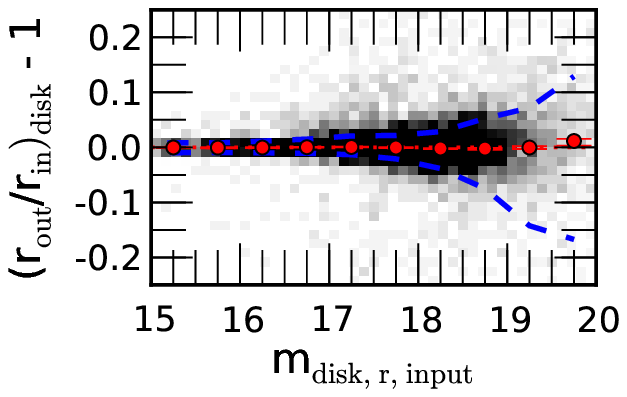}
\includegraphics[width=0.99\linewidth]
{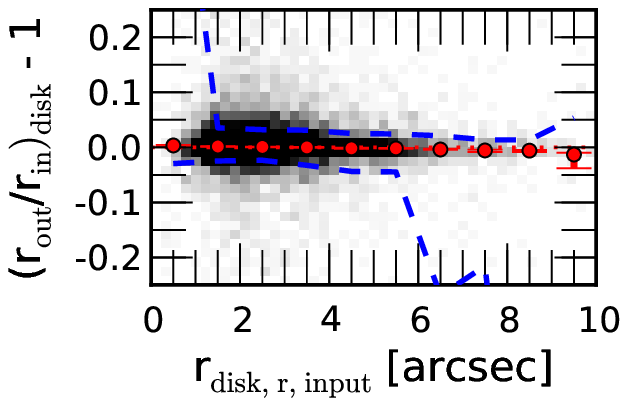}
\subcaption{}\label{subfig:pix2_serexp_serexp_2com}
\end{minipage}

\caption{The simulated and recovered
component parameters for a \SerExp\ galaxy 
fit with a \SerExp\ model in four cases:
\textbf{\subref{subfig:flat_serexp_ser}}
the image prior to adding Poisson noise, \textbf{\subref{subfig:psf_serexp_ser}}
our fiducial case containing simulated sky, Poisson noise, PSF errors, and neighboring
sources,
\textbf{\subref{subfig:sn4_serexp_ser}} the fiducial case with S/N increased by
a factor of 4, and \textbf{\subref{subfig:pix2_serexp_ser}} the fiducial case
with resolution increased by a factor of 2. 
Over-plotted are the bias (red points) in the fitted values. 
All plots show the 68\% (dashed line) scatter in blue. The density
of points is plotted in gray-scale.
}
\label{fig:serexp_serexp_2com}

\end{figure*}

The total apparent magnitude, halflight radius, and additional fit
parameters recovered by fitting the noiseless images of the \Ser\
and \SerExp\ models are presented in Figures \ref{subfig:flat_ser_ser},
\ref{subfig:flat_ser_serexp}, \ref{subfig:flat_serexp_ser},
\ref{subfig:flat_serexp_serexp}, and \ref{subfig:flat_serexp_serexp_2com}. The
plots show the difference 
in simulated and fitted values (fitted value - input value). The difference is
shown versus the input magnitude as well as the input value of the respective
fit parameter. The gray-scale shows the density of points in each
plane with red points showing the median value. Error bars on the median value
are the 95\% confidence interval on the median obtained from bootstrapping.  
Blue dashed lines show the regions which contain 
68\% of the objects.

Figures \ref{subfig:flat_ser_ser} and \ref{subfig:flat_serexp_serexp} show the
corresponding fit is well constrained (\Ser\ fit with \Ser,
and \SerExp\ with \SerExp). The total magnitude and halflight
radius are both constrained well within 1\% error on the flux or radius
($\sigma_{\mathrm{total\ mag}} \approx 0.01$ mag and
$\sigma_{\mathrm{radius}} \approx 1\%$). However, the scatter increases
somewhat for the sub-components of the \SerExp\ fit (see Figure
\ref{subfig:flat_serexp_serexp_2com}). As the
components of the \SerExp\ model become dim (bulge/disk magnitude approaches
18.5), the component contribution to the total light becomes small. The
origin of the magnitude limit is merely an artifact of our selection criteria
requiring that all galaxies have total magnitude brighter than 17.77. This
implies that components with magnitude of $\approx$18.5 or dimmer are
necessarily sub-dominant
components and contribute at most $\approx$50\% of the light to the total profile. On
average, components dimmer than 18.5 magnitudes contribute about 25\% of the
total light to a typical galaxy in this sample, and this contribution drops
rapidly to about 10\% by 19 magnitudes. In these
cases, the sub-dominant component will be much less apparent in the image and,
therefore, less important to the overall $\chi^2$ of the fit, allowing for
greater error in the parameters of that component. In addition, once
Poisson noise is considered, these dimmer components suffer from much
lower S/N. Later tests (Section~\ref{subsec:sn_psf}) show substantial error on
these components due to the low flux and resulting low S/N.

Additionally, sub-dominant components (in particular, bulges) may be much
smaller than the overall size of the galaxy. This makes bulge parameter
recovery susceptible to resolution effects. These effects are also explored in
Section~\ref{subsec:sn_psf}.
 
The magnitude and halflight radius are also well constrained when a \Ser\ galaxy
is fit with a \SerExp\ profile (Figure \ref{subfig:flat_ser_serexp}). However, a
\SerExp\ galaxy fit with a \Ser\ profile produces large biases in the magnitude
and halflight radius (Figure \ref{subfig:flat_serexp_ser}). 

As already mentioned, the total magnitude and halflight radius are well
constrained 
($\sigma_{\mathrm{total\ mag}} \approx 0.01$ mag and
$\sigma_{\mathrm{radius}} \approx 1\%$) in cases where the correct model
is applied to the mocks (\ie \Ser\ mock fit with a \Ser\
model). This is not always the case when the wrong model is applied (\ie
\SerExp\ mock fit with a \Ser\ model). When attempting to fit the
simulated \SerExp\ mocks with a \Ser\ model, we find measurable bias of
order .01 magnitudes in total magnitude. We also find the scatter of both the
size and magnitude to be increased by an order of magnitude. This bias and
increased scatter becomes even larger in later tests. It is obvious that a
single-component galaxy cannot properly model a two-component
galaxy in general, and therefore, we would expect significant problems in 
attempting to fit a single-component profile to a two-component galaxy. 
Nevertheless, this type of fit is often performed on real data at low to
moderate resolution and S/N where it is unlikely to 
recover a robust two-component fit. An important observation is that the
\SerExp\ fit provides the most stable estimate of the halflight radius and total
magnitude regardless of the true simulated galaxy model (\Ser, \DevExp, or
\SerExp). The additional freedom in the \SerExp\ model and the 
fact that the \Ser\ and \DevExp\ models are special cases of the 
\SerExp\ model would lead us to expect this result. Therefore, it is
advisable to always use a \SerExp\ fit in the course of fitting SDSS galaxies
unless there is specific evidence to the contrary.

One systematic effect in the pipeline that has been noted by other groups
\citep[\eg][]{nyu_vagc, guo2009}, is the underestimate of S{\'e}rsic  index at
larger S{\'e}rsic  indexes. At S{\'e}rsic  indexes of $n\approx$ 4, we
underestimate the S{\'e}rsic  index by less 
than $1\%$. However, this underestimate increases in the later
tests. The data suggest that a substantial component of this error is due to the
resolution limits of the SDSS sample. At larger S{\'e}rsic index, a high
sampling rate at the center of the galaxy is useful in distinguishing the
preferred value of the S{\'e}rsic index. We further explore the effect of image
resolution in Section~\ref{subsec:sn_psf}. 

Since no Poisson noise is added to these images, the scatter
apparent in these fits is a combination of the limitations of the SDSS data (in
particular resolution), systematics inherent in the PyMorph routine (as well as
the GALFIT routine used by PyMorph), and any parameter degeneracies inherent in
the models.

GALFIT uses the Levenberg-Marquardt minimization method \citep{Crecipies} 
to find the minimum of the $\chi^2$ distribution of the fit. 
The Levenberg-Marquardt method is not a global search algorithm 
but rather follows the steepest decent to a local minimum.
As the parameter space becomes more complicated, 
GALFIT has more trouble accurately recovering parameters. 
Adding components to the fit (\ie going from a one-component to two-component
fit or going from a fixed S{\'e}rsic  index component to one with a free S{\'e}rsic 
index) will not only complicate the $\chi^2$ surface, making convergence less
likely, but may introduce true degeneracies in the parameter space. 

For instance, the \SerExp\ fit of a galaxy of very late type often suffers from
over-fitting. The bulge component will tend to fit the disk of the galaxy as a
second disk component with $n_{\mathrm{bulge}}\approx1$. This is obviously
an unintended solution to the fitting but one that is equally valid
from an $\chi^2$ perspective. In practice, it is difficult to prevent
this type of convergence without artificially constraining the fitting routine.
Such constraints are generally discouraged and can lead to other negative effects
including convergence to a non-optimal solution. The best
solution to the parameter degeneracy is close examination of any two-component
fits in cases where $n_{\mathrm{bulge}}\approx 1$, or B/T $\approx 0$ or 1.

In addition, PyMorph reports statistical error estimates on the
fitted parameters as returned from GALFIT. These errors are found to be an
underestimate of the true error in the fits by as much as an order of magnitude.
This gross underestimation of the error is also reported by \cite{GEMS2007} as
well as being discussed in the GALFIT user notes\footnote{See the technical
FAQs at \url{ http://users.obs.carnegiescience.edu/peng/work/galfit/TFAQ.html}}.
Following \cite{GEMS2007}, we examine the ratio of the
uncertainty reported by GALFIT to the deviation of the measured parameters
(referred to as $\sigma/\Delta$). $\sigma/\Delta$ should be greater than 1
for approximately 68\% of the data if the estimated uncertainty is appropriate.
However, this is not the case for any of the parameters in the fits. 
We discuss a simple method for correcting the
systematic bias and estimating the uncertainty in Section~\ref{sec:discuss}.
 
\subsection{The effects of background, neighbor sources, and incorrect PSF
extraction} \label{subsec:psf_test}
When analyzing real data, it is not possible to extract the PSF at the 
target galaxy to arbitrary accuracy. Interpolation is required and generally
performed on a network of the nearest stars to the target galaxy. 
We test this effect through extraction of a neighboring PSF to
be used during fitting in place of the PSF used to generate the image. 

The neighbor PSF used in fitting is randomly selected from a location within a
200 pixel box surrounding the source. This provides approximately even
sampling of distances from nearly 0 to about 170 pixels in separation from the
source which corresponds to a separation of $\approx0$ to $\approx67.32$
arcseconds
between the target galaxy and the location used for PSF extraction. 
This inserts some PSF error into the process of fitting as would be expected 
in the case of real data. However, it also retains the similarity between the PSF
used for simulation and the PSF used for fitting. A strong similarity between 
the two would be expected since the PSF generally will not vary greatly over 
the area of a single fpC image.

Target galaxies are randomly inserted into the 
simulated fpC images described in Section~\ref{subsec:background}.
The simulated fpC images contain sky as well as neighboring sources.
The PSF of the neighboring sources will
have a different PSF than the target galaxy. This effect is not of concern in
this work. 

Prior to fitting, a new cutout is extracted from the
total image (containing the target galaxy and background) ensuring that the
target galaxy is
at the center of the stamp image. By constructing new postage stamp images in
this manner, we ensure that there is sufficient variation in the background
while preventing us from fitting the incorrect galaxy. 

These fits (containing error in PSF reconstruction, neighboring sources, and
noise) are the closest simulation to actual observing conditions that
we have analyzed. Therefore the fits and the resulting measures of scatter and
bias are adopted as our fiducial estimates of scatter and bias when using the
pipeline. 

Figures \ref{subfig:psf_ser_ser}, \ref{subfig:psf_ser_serexp},
\ref{subfig:psf_serexp_ser}, \ref{subfig:psf_serexp_serexp}, and
\ref{subfig:psf_serexp_serexp_2com}
show that we recover the input values with marginal scatter. 
The total magnitude and halflight radius remain well constrained 
($\sigma_{\mathrm{total mag}} \approx 0.05$ mag and
$\sigma_{\mathrm{radius}} \approx 5\%$) in cases where the 
correct model is fit to the mock galaxy. However, this scatter 
becomes larger when the wrong model is fit. The underestimate of 
the S{\'e}rsic  index, particularly at large values, persists.

Further examination of the two-component fits show that the pipeline has 
difficulty extracting dim components (bulge or disk magnitude dimmer than 
$\approx 18.5$). In these ranges, the components are observed at lower S/N
and the pipeline looses sensitivity to the model parameters. 
The \SerExp\ fit shows an  underestimate of S{\'e}rsic  index, which is even
stronger than in the single-component case, and an underestimate of 
bulge radius.  However, the disk parameters remain unbiased with an 
increase in scatter of the model parameters. The increased stability of
the disk parameters relative to the bulge parameters was also noted in
\cite{simard11}. In their paper, the authors comment that this may be due to the
fixed profile shape (due to the fixed S{\'e}rsic index, $n=1$) or to the fact that
on average bulges are more compact than disks leading to a resolution effect.
This stability is the result of the increased resolution as disk sizes 
in our
sample are roughly 3 times the FWHM of the PSF while bulges are smaller, on average
approximately equal to the FWHM of the PSF in size. We discuss this further in
Section~\ref{subsec:sn_psf}.

In general, the \SerExp\ fits are problematic and require much care when
analyzing individual components. However, as we have already shown,
total magnitude and halflight radius are still tightly constrained. 

\begin{figure}
\includegraphics[width=\columnwidth]{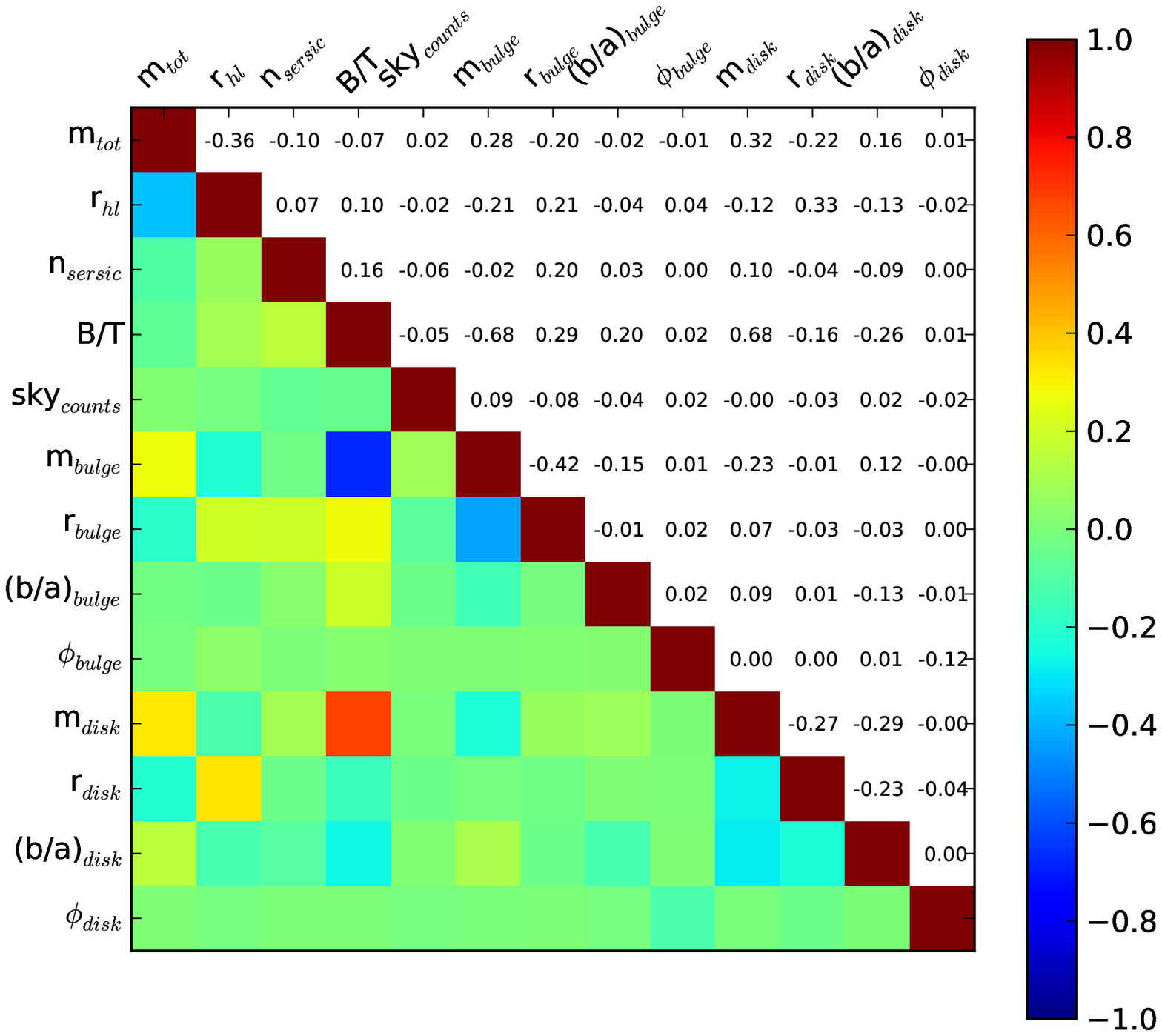}
 \caption{The correlation matrix for a mock \SerExp\ galaxy fit with a
\SerExp\ model.}
\label{fig:correlation_serexp_serexp}
\end{figure}

Table \ref{table:psf} summarizes the bias and scatter in the fits;
they exhibit trends with both the input value of the parameter
and the input magnitude of the galaxy. This behavior is not properly
encapsulated in the overall measure of bias, so these values are useful only 
as an example of the relative scale of bias and scatter for each parameter.

Errors can be correlated across many fit parameters, so we
also calculate a correlation matrix for the parameter errors. Figure
\ref{fig:correlation_serexp_serexp} shows an example of the
correlation matrix for the \SerExp\ mocks fit with a \SerExp\
model. We see the expected strong correlations between
bulge-to-light ratio and the bulge and disk magnitudes as well as the
correlation among the radii of the bulge component with the S{\'e}rsic index.
While the correlation matrix suggests that there is little correlation
between sky estimation error and the fitted parameters, we will show later that
there is indeed a strong correlation in model errors with sky estimation error. 

The apparent lack of correlation of sky error with the other
fitting parameters is somewhat surprising. However,
Figures~\ref{fig:bkrd_ser_ser} and \ref{fig:bkrd_serexp_serexp} suggest a
possible explanation for the apparent lack of correlation. Correlation of
parameter errors with sky errors is non-linear and asymmetric with respect to
over- or underestimating the sky. The fits discussed in this section are shown on
Figures~\ref{fig:bkrd_ser_ser} and \ref{fig:bkrd_serexp_serexp} in red. These
points lie in a region where sky error does not significantly bias most
parameters. In addition, the scatter of the sky values is quite small. This
small scatter prevents us from sampling the broader covariance of the sky. If,
for example, the recovered sky value
was an underestimate of 0.5\%, then there would be a measurable covariance of
fitting parameters with sky due to the steepness of the parameter bias with respect
to sky level. We discuss the sky estimation further in
Section~\ref{subsec:bkrd_offset}.

\begin{table*}
\begin{tabular}{c c c c c c }
    Simulated model  & Parameter & \multicolumn{4}{c}{fitted model}\\ \cmidrule{3-6}
     & & \multicolumn{2}{c}{\texttt{Ser}} &
    \multicolumn{2}{c}{\texttt{SerExp}} \\ \cmidrule{3-4} \cmidrule{5-6} 
      &  & bias & 1-$\sigma$ & bias & 1-$\sigma$ \\
    \hline 
\multirow{4}{*}{\texttt{Ser}}  & m$_{tot}$ [mag]  & $0.00\pm 0.09$ & [$-0.02,
0.02$]  & $-0.02\pm 0.18$ & [$-0.11, 0.02$]  \\
 & r$_{hl}$ [arcsec]  & $-0.01\pm 0.68$ & [$-0.14, 0.01$]  & $0.11\pm 1.53$ & [$-0.11, 0.04$]  \\
 & sky [\%]  & $-0.05\pm 0.14$ & [$-0.11, -0.05$]  & $-0.05\pm 0.14$ & [$-0.11, -0.05$]  \\
 & S{\'e}rsic Index  & $-0.08\pm 0.57$ & [$-0.44, 0.01$]  & -- & --  \\
\hline 
\multirow{9}{*}{\texttt{SerExp}}  & m$_{tot}$ [mag]  & $-0.07\pm 0.18$ &
[$-0.22, -0.01$]  & $-0.02\pm 0.15$ & [$-0.09, 0.01$]  \\
 & r$_{hl}$ [arcsec]  & $0.49\pm 1.86$ & [$-0.02, 0.30$]  & $0.07\pm 1.17$ & [$-0.07, 0.04$]  \\
 & sky [\%]  & $-0.08\pm 0.15$ & [$-0.16, -0.07$]  & $-0.06\pm 0.13$ & [$-0.11, -0.05$]  \\
 & B/T  & -- & --  & $0.00\pm 0.15$ & [$-0.07, 0.03$]  \\
 & m$_{bulge}$ [mag]  & -- & --  & $-0.14\pm 0.71$ & [$-0.73, 0.06$]  \\
 & m$_{disk}$ [mag]  & -- & --  & $-0.04\pm 0.50$ & [$-0.41, 0.05$]  \\
 & r$_{bulge}$ [arcsec]  & -- & --  & $0.08\pm 0.97$ & [$-0.27, 0.11$]  \\
 & r$_{disk}$ [arcsec]  & -- & --  & $0.07\pm 0.82$ & [$-0.14, 0.08$]  \\
 & S{\'e}rsic Index & -- & --  & $0.06\pm 1.98$ & [$-0.90, 0.14$]  \\
\end{tabular}

\caption{The bias and scatter of the fitted parameters of the simulated 
images with background and PSF effects. These values are provided for 
illustrative purposes
only. There is much underlying structure in the errors when compared to their
respective input values or the magnitude of the component.}
\label{table:psf}
\end{table*}

\subsection{Testing with real images} \label{subsec:real}
To verify the validity of the simulated background and to test the fitting
pipeline
in clustered environments, we insert the mock galaxies
into real SDSS fpC images. The fpC images are selected from SDSS DR7 images
containing spectroscopic galaxy targets.

We omit plots of the fitted values here, because the scatter and
the bias in the fits remain unchanged, suggesting that we have properly modeled
the sky background and neighboring sources common to an SDSS spectroscopic
galaxy.

Dense environments provide an additional test for our pipeline. 
To select fpC images that contain dense environments, we use the GMBCG 
catalog \citep{GMBCG}. We match brightest cluster galaxies (BCGs) with
galaxies in our original 
catalog to select fpC images with cluster members including the BCG. 
Our mock galaxies are then inserted into the image which is run through the
pipeline. 
In our previous simulations, intracluster light and gradients in the 
sky were not modeled. These tests allow us to see what the effects may be. 
Once again, the errors remain unchanged, showing that no environmental 
correction is necessary when using the fits from the pipeline. 

Placing mock galaxies near cluster members allows us to test for 
systematic effects in crowded fields. However, further examination 
of BCG galaxies is necessary before we are able to properly model them 
for this purpose. For example, the curvature at the bright end observed 
in the size-luminosity relation of early-type galaxies \citep[see
][]{Bernardi2013} appears to be due to 
an increasing incidence of BCGs, which define steeper 
relations than the bulk of the early-type population 
\citep[\eg][]{BernardiHyde2007, Bernardi2013}. However, the curvature could
also be due to intracluster light \citep[\eg][]{Bernardi2009}. Our
ability to test the systematic effects associated with BCGs using the method
outlined
above is severely limited due to the existence of a BCG at the location we would
prefer to place our test galaxy (\ie the center of the cluster).
Therefore, the stability of recovered fit parameters with respect to
environment cannot be assumed to extend to BCGs based on the tests presented
here alone. Further tests for the largest, brightest galaxies are needed to
explore this possibility. We have not presented these tests in this text.

\subsection{Varying the S/N and pixel size} \label{subsec:sn_psf}
In addition to the previous tests, we isolate the effects of the S/N and image
resolution to quantify the contributions to the bias and scatter in
our
fits. Figures \ref{subfig:sn4_ser_ser}, \ref{subfig:sn4_ser_serexp},
\ref{subfig:sn4_serexp_ser}, \ref{subfig:sn4_serexp_serexp}, and
\ref{subfig:sn4_serexp_serexp_2com}
 show the effect of increasing the S/N by a factor of 4 while holding all other
parameters fixed. Similarly, Figures \ref{subfig:pix2_ser_ser},
\ref{subfig:pix2_ser_serexp},
\ref{subfig:pix2_serexp_ser}, \ref{subfig:pix2_serexp_serexp}, and
\ref{subfig:pix2_serexp_serexp_2com} show the effect of increasing
resolution by a factor of 2 while holding S/N constant. Corresponding
decrements in these parameters were performed, although they are not presented
in this paper. 

Improving the resolution by a factor of two substantially improves the
ability to recover the radius and S{\'e}rsic index with reduced bias. For
instance, the S{\'e}rsic index bias is reduced to $\approx0.1$ at the larger
values. Additionally, the bulge parameters of the \SerExp\ fit
improve substantially with improved resolution.
Corresponding changes in the S/N reduce the scatter, but by a small amount
relative to the effect of the resolution
change. In addition, changing the S/N does not remove the observed bias in
S{\'e}rsic index or bulge size. This leads us to conclude that the limitations
of the resolution of SDSS are the leading factor
in causing systematic offsets in the halflight radius, S{\'e}rsic index, and
other fitting parameters (including the bulges of the \SerExp\ fits). While
increasing the S/N will reduce the scatter in the fits, increased resolution is
necessary to properly recover unbiased values. 

\cite{Lackner2012} also examined the effects of changing S/N and
resolution on SDSS galaxies (see Figures 5-11 of their paper). The authors
found that decreased resolution and S/N increases the relative error in the
S{\'e}rsic index and radius. They recommended that \Ser\ galaxies (and
the bulge and disk sub-components of two-component galaxies) have radii,
$R_{hl}\gtrsim0.5\times$FWHM.  This cut removes $\approx1\%$ of the
\Ser\ mocks and $\approx 22\%$ of the \SerExp\ mocks from our simulated samples
with a preference toward galaxies above $z=0.05$.

While this condition is sufficient to keep the relative error in the halflight
radius and S{\'e}rsic index comparable to the error in the magnitude, we find
that this condition fails to remove the bias in our galaxy samples.
Figure~\ref{subfig:psf_ser_ser} shows that
the underestimate of S{\'e}rsic index occurs at larger values. These galaxies
tend to exhibit radii larger than the PSF. Given that the average FWHM of PSFs in our sample
is $\approx1.3''$, if we apply the suggested cut in radius, we are unable to
remove the bias in S{\'e}rsic index. Clearly, reliable measurements
are dependent on both the S{\'e}rsic index of the object and its radius
relative to the resolution. Both parameters must be accounted for when
deciding on an appropriate resolution cut.  

If we extend the \cite{Lackner2012} recommendation to include a
S{\'e}rsic index dependent term, this is sufficient to provide for recovery of
S{\'e}rsic index $>$ 4 with bias $\approx 0.1$ or $\approx1\%$. Galaxies should have
circularized halflight radii $R_{hl}\gtrsim$0.5*FWHM$\times n$. This
removes nearly 75\% of the sample. While such large
cuts are sufficient to remove the bias in radius and S{\'e}rsic index for the
\Ser\ fits, the data are certainly biased relative to our original catalog after
the cuts. Rather than remove
these galaxies, we correct for the bias following a simple statistical argument
presented in Section~\ref{sec:discuss}.

\subsection{Effect of cutout size}\label{sec:cut_size}
We select postage stamp cutouts for use in fitting. It is important to select a
cutout size that does not significantly bias the fits produced by PyMorph. 
The most important consideration is to provide enough sky pixels to allow 
the PyMorph program to properly determine the sky level in the images. 
Reducing cutout size may cause overestimation of background and corresponding 
errors in the other fit parameters. 
However, we use the PyMorph pipeline and GALFIT to fit a constant background
to the galaxy image. Since a larger image could make sky gradients more 
significant, this could bias the fits when a larger cutout is used. 
We seek to minimize error when estimating the sky level without introducing a
gradient term and further complicating the fitting process. 

\begin{figure*}    
\includegraphics[width
=0.30\linewidth]{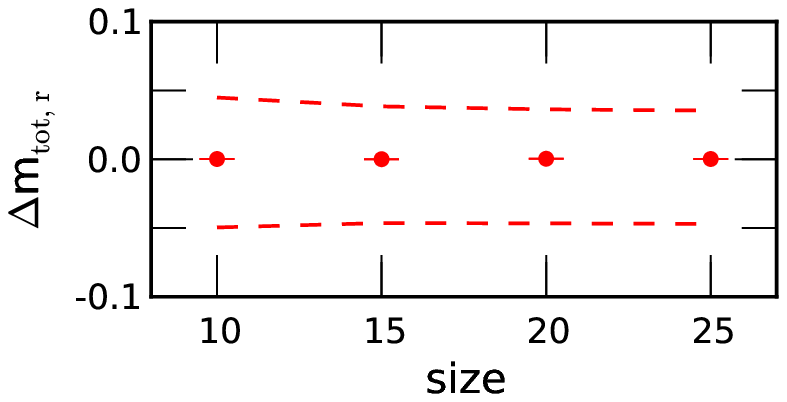}
\includegraphics[width =
0.30\linewidth]{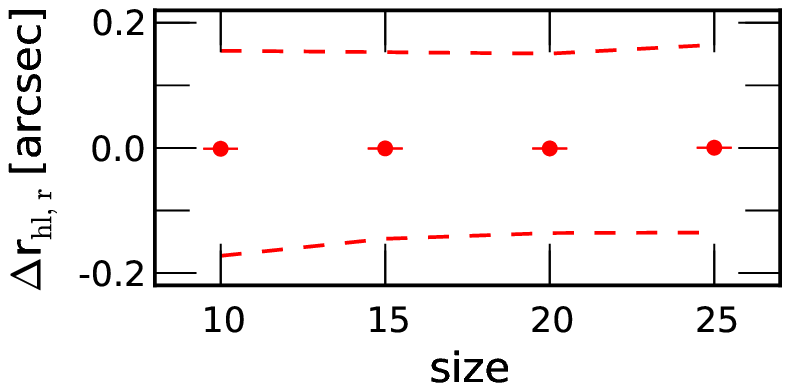}
\includegraphics[width =
0.30\linewidth]{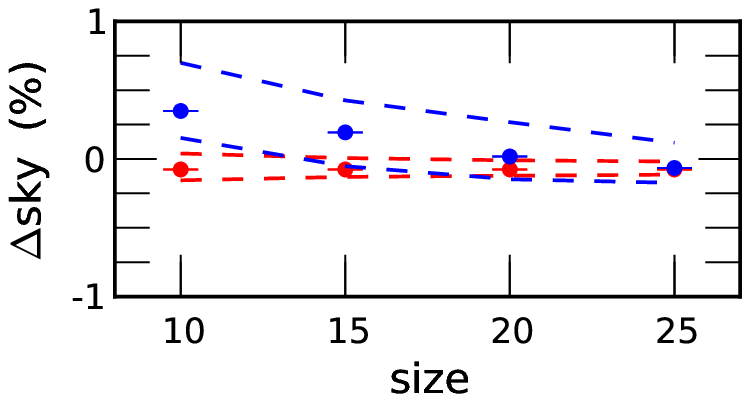}
\caption{The mean difference of the total magnitude
(left column), PSF-corrected halflight radius (center column), and sky estimation
(right column) as a function of cutout size for \SerExp\ mocks 
fitted with a \SerExp\ model. Other simulated models behave similarly. 
For sky estimation, the sky measured by GALFIT is plotted
in red. SExtractor sky measurements are plotted in blue for reference.  
One-$\sigma$ scatter in the fits is plotted as a dashed line. 
Improvement in scatter when fitting for cutout sizes above 20 Petrosian radii is
limited, so we use a 20 halflight radii cutout size for all images. 
Fit parameters seem to have no sensitivity to cutout size in this range, 
suggesting that it may even be possible to use smaller cutouts.}
\label{fig:cutout_size_all}
\end{figure*}

To test for optimal cutout size, we fit mocks with cutout sizes
between 10 and 25 Petrosian half-light radii. We plot the average difference
between simulated and measured fit parameters below. 
In Figure \ref{fig:cutout_size_all} we present the error and 1-$\sigma$ 
scatter in the error on the total magnitude, halflight radius and sky 
(showing SExtractor sky in blue and our estimates in red) as a function 
of cutout size. 
Smaller sizes clearly bias sky estimates made by SExtractor, but only minor
improvement in the scatter of any parameters is achieved by using cutout sizes 
above 20 halflight radii. Since we use SExtractor sky as a starting point for
our fitting, we choose a size of 20 halflight radii for our images. The sky
estimates of SExtractor improve substantially. However, GALFIT sky
estimation is stable over these sizes. Because GALFIT sky estimation is 
largely independent of the initial starting SExtractor value (which we 
would expect if we are truly finding the best fit to the galaxy), 
it is likely the case that cutout sizes smaller than even 10 halflight 
radii could be used for analysis. 

Additional plots of other parameters are omitted in this section. 
The other fitted parameters show little or no sensitivity to cutout size 
in the range of cutout sizes used. However, as previously discussed, the 
bias and scatter may not be equally affected across all model parameters.
The effects may be concentrated in a small part of the parameter space.  

\subsection{The effect of incorrect sky estimation}\label{subsec:bkrd_offset}
Estimation of the sky in the vicinity of the target galaxy has a high level of
uncertainty. Indeed, accurate sky determination is likely not even a solvable
problem as discussed briefly in \cite{Blanton2011}. 
To determine the bias introduced by our sky estimation, we have 
tested our fitting pipeline in cases of both underestimation and 
overestimation of the sky. We fix the sky at the simulated sky level, 
as well as at simulated sky level $\pm0.5\%$ and $\pm1.0\%$. These
ranges were chosen to represent the range of differences between our sky 
estimations and those provided in the CASJOBS database for the SDSS 
spectroscopic sample.

\begin{figure}
\centering
\includegraphics[width =\linewidth]{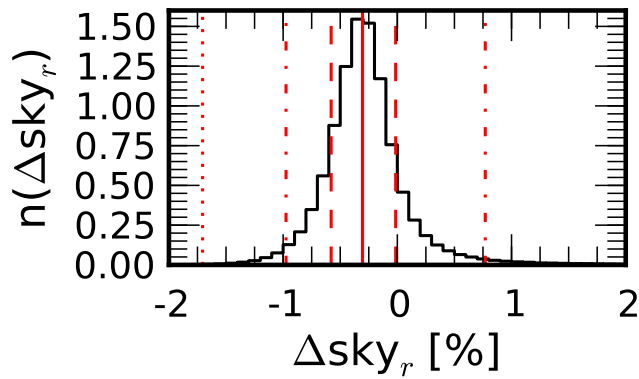}
\caption{The percent difference between the sky estimate of PyMorph for SDSS
galaxies and the sky estimated by the SDSS photometric pipeline for those same
galaxies.
The normalized distribution of differences is shown in bins of 0.1\%. A negative
difference indicates that the sky measured by PyMorph is lower than that
reported by SDSS. The vertical red solid line indicates the median of the
distribution. 
The red dashed, dot-dashed, and dotted lines indicate the 68-95-99\% ranges of the
data, respectively.
The 95\% range of sky values is approximately between $\pm1\%$ difference, so we
adopt this as the range used to test the effects of improper sky estimation.}
\label{fig:sky_diff}
\end{figure}

Figure \ref{fig:sky_diff} shows a comparison of sky estimates using PyMorph to
those provided from the SDSS photometric data pipeline. 
This comparison is performed on data from the catalog presented in \catalog.  
The Figure shows the normalized distribution of differences in sky estimation 
in bins of 0.1\%. A negative difference indicates that the sky measured 
by PyMorph is lower than that reported by SDSS. 
The vertical red solid line indicates the median of the distribution. 
The red dashed, dot-dashed, and dotted lines indicate the 68-95-99\% ranges 
of the data, respectively. The 95\% range of sky values is approximately between
$\pm1\%$
difference. 
For the test, we adopt this range as the range to test for sky variation.

\begin{figure*}
\includegraphics[width=0.30\linewidth]
{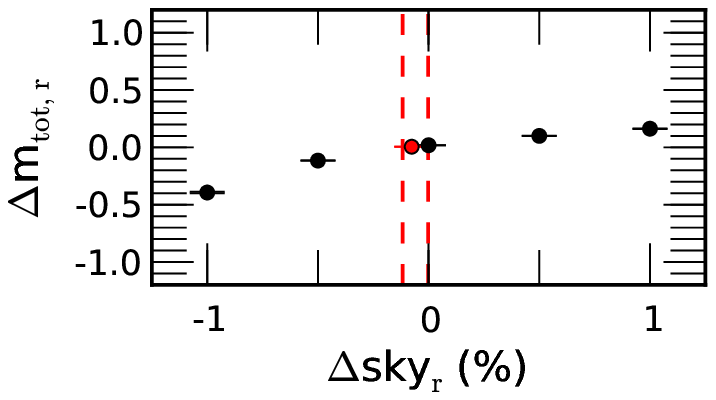}
\includegraphics[width=0.30\linewidth]
{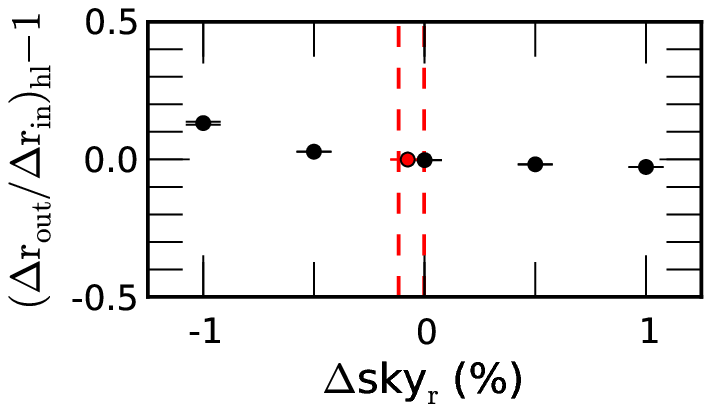}
\includegraphics[width
=0.30\linewidth]{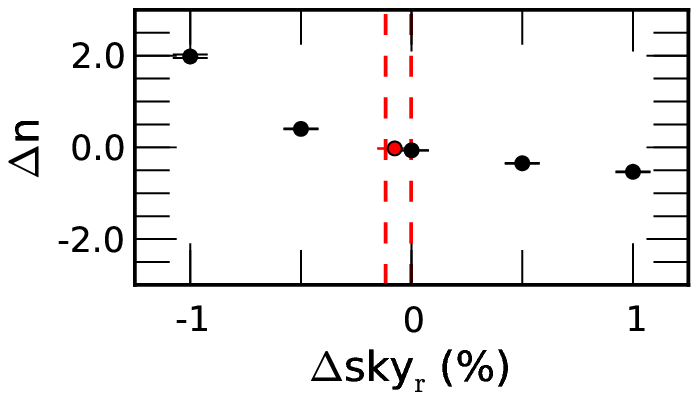}

\caption{The simulated and recovered apparent magnitude (left), halflight
radius (center), and S{\'e}rsic index (right) for a \Ser\ galaxy 
fit with a \Ser\ model
The residuals are plotted as a function of the sky level. Points plotted in
black are from fits performed with fixed sky. The overplotted points in red are 
the result of fitting with sky level as a free parameter in the fit. The
vertical dashed red lines mark the 68\% scatter of the free sky determination.
Our fits are slightly biased low, and this contributes to a small overall bias
in fit parameters.}
\label{fig:bkrd_ser_ser}
\end{figure*}

\begin{figure*}

\includegraphics[width=0.30\linewidth]
{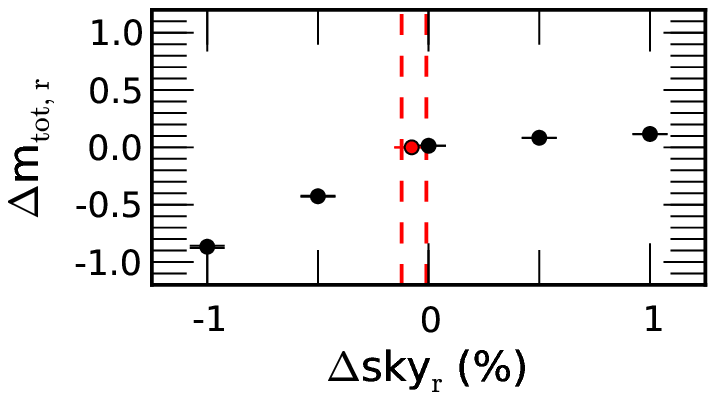}
\includegraphics[width=0.30\linewidth]
{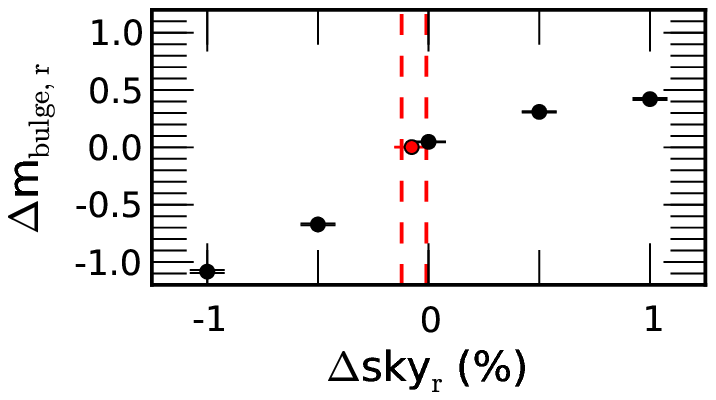}
\includegraphics[width=0.30\linewidth]
{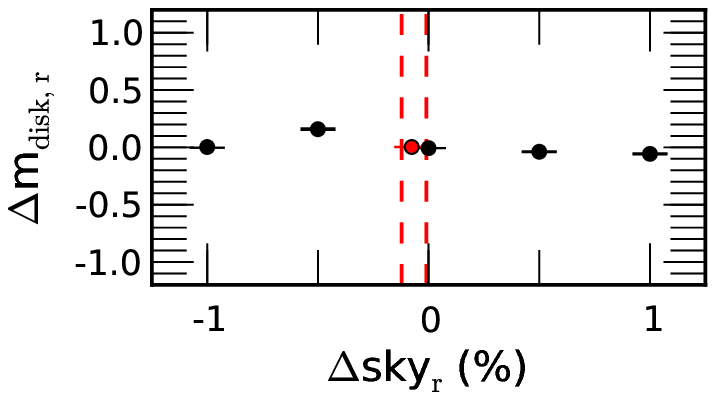}

\includegraphics[width=0.30\linewidth]
{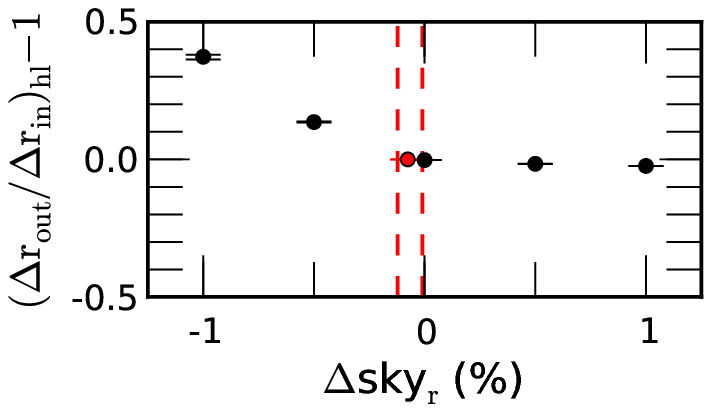}
\includegraphics[width=0.30\linewidth]
{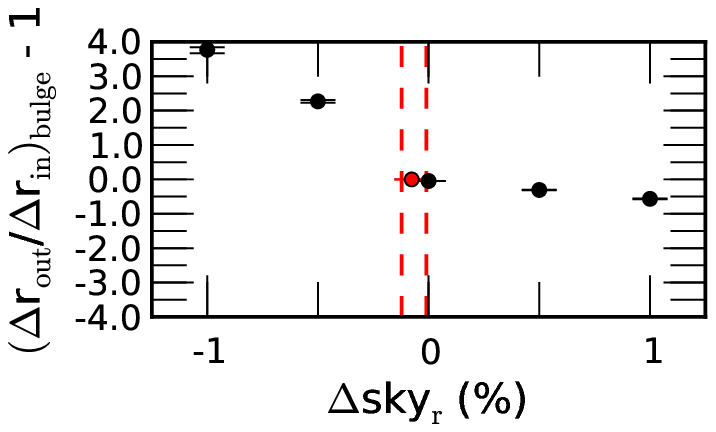}
\includegraphics[width=0.30\linewidth]
{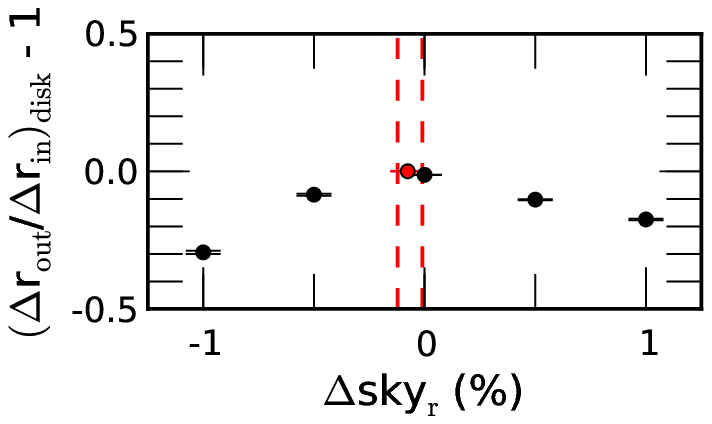}

\includegraphics[width=0.30\linewidth]
{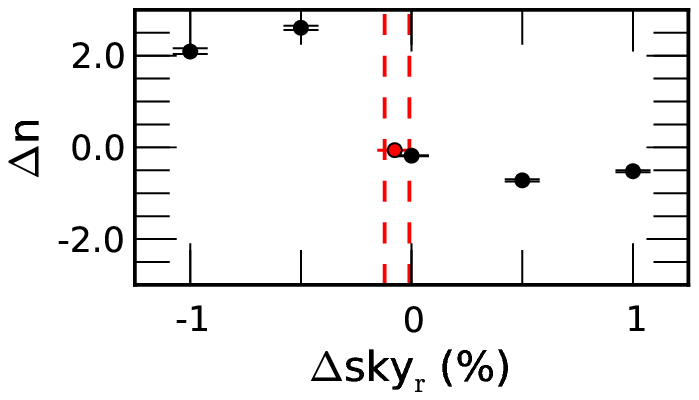}
\includegraphics[width=0.30\linewidth]
{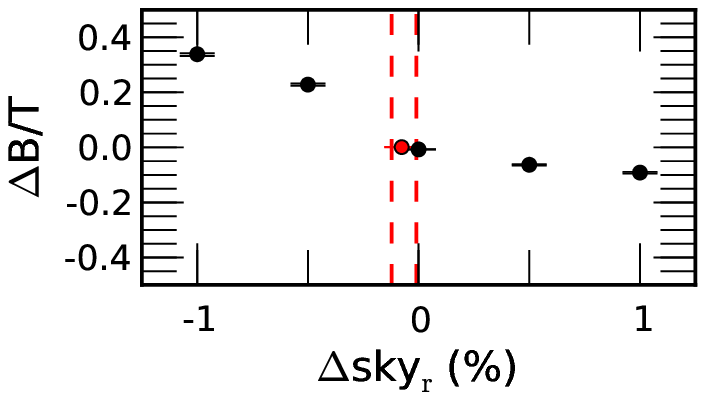}
\includegraphics[width=0.30\linewidth]
{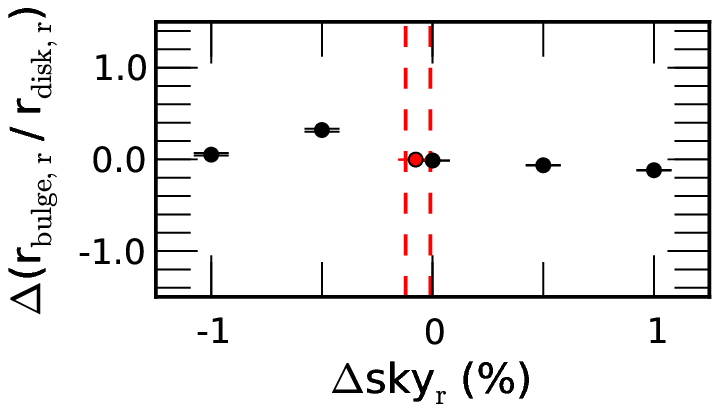}

\caption{The simulated and recovered parameters of a simulated \SerExp\ galaxy
fit with a \SerExp\ profile.
The residuals are plotted as a function of the sky level. Points plotted in
black
are from fits performed with fixed sky. The overplotted points in red are 
the result of fitting with sky as a free parameter of the model. The
vertical
dashed red lines mark the 68\% scatter of the GALFIT sky determination. Notice
that disk parameters are relatively robust while bulge parameters suffer from
incorrect sky estimation. Underestimates of sky level have a particularly
strong effect on the bulge.
}
\label{fig:bkrd_serexp_serexp}

\end{figure*}

The results of incorrectly estimating the sky are shown in Figures
\ref{fig:bkrd_ser_ser} and \ref{fig:bkrd_serexp_serexp}. 
In red, we show the results of fitting galaxies using
the standard PyMorph pipeline, treating sky level as a free parameter in the fit.
PyMorph systematically underestimates the sky at the 0.1\% level. However, the
scatter is very tight as indicated by the vertical dashed red lines. 
In black we have plotted the fitting results at fixed sky levels
of the correct value and $\pm 0.5\%$ and $\pm 1.0\%$. Errors approaching
0.5\% clearly introduce a large bias in the fits. The
0.5\% level is an important level because it is the
approximate level of overestimation shown in the preceding section
(Section~\ref{sec:cut_size}) found by SExtractor.  

Note the asymmetry of the effects of incorrect sky estimation on fitting 
parameters. In particular, an underestimate of sky is much more
detrimental to the fit than the corresponding overestimate. 
The reason for this asymmetry is due to changes in the perceived
``flatness'' of the profile at large radii. When the sky is overestimated, the
galaxy profile tends to 0 flux too early. This causes a decrease in the S{\'e}rsic
index and a decrease in the radius.  However, when the sky is underestimated,
there will be an extended, approximately constant brightness profile at larger 
radii. The only way to model such a profile is for S{\'e}rsic index to diverge to
larger values which produce flat, extended profiles at large radii.

\cite{guo2009} examined the effects of sky uncertainties in regards to the
covariance between magnitude and both S{\'e}rsic index and halflight radius.
They randomly sampled sky estimates from a distribution contained mostly within
$\pm1\%$. 
They found similar variation of S{\'e}rsic index (varying by 2 or more in some
cases of underestimating the sky and varying by less than 1 in the case of
overestimation).   
The asymmetry in bias due to incorrect sky estimation is apparent in Figure 5
of \cite{guo2009}, but not explicitly commented upon. 

 Figure \ref{fig:sky_diff} shows that PyMorph consistently 
estimates the sky $\approx$0.25\% lower than that of the SDSS pipeline.
Figures~\ref{fig:cutout_size_all}, \ref{fig:bkrd_ser_ser}, and
\ref{fig:bkrd_serexp_serexp} show that PyMorph has a systematic
underestimate of the true sky at the $\approx$0.1\% level. 
This bias is much smaller than the bias associated with using the SExtractor sky
estimate as shown in Figure~\ref{fig:sky_diff} (especially for smaller cutout 
size), which suggests that the sky 
values in SDSS are slightly overestimated.

\SerExp\ disk components are remarkably robust to the errors in sky estimation,
while bulge parameters suffer greatly, especially when the sky is
underestimated. Upon further examination of Figure
\ref{fig:bkrd_serexp_serexp}, 
the bulge parameters of the model are more accurately estimated when the sky
is treated as a free parameter in the fit rather than when the sky is fixed at the
correct value. However, this improvement does not suggest that underestimate of the
sky is the
preferred fitting outcome. It merely reflects the fact that the systematic
effects due to underestimation of the sky are opposite to the underlying biases
in halflight
radius and S{\'e}rsic index. If we were to apply the PyMorph pipeline to an
image
with higher S/N and increased resolution, we would prefer the correct estimate of
the sky to prevent systematic overestimate of these parameters.
  
\section{Discussion}\label{sec:discuss}
In the preceding sections we have shown the covariance, bias, and 
scatter in our parameter estimation for the \Ser\ and \SerExp\
models. In reality, the effects above will combine to yield a total
scatter, covariance, and bias that should approach those shown in
\ref{subsec:psf_test}. Our simulations give us an idea of the behavior 
of the PyMorph pipeline when fitting SDSS galaxies as presented in \catalog. 

The simulations show that the recovery of global fitting parameters 
(total magnitude and halflight radius) in the case of SDSS galaxies is 
remarkably robust, even in the case of the \SerExp\ fits. Two-component fits
present a more difficult test for the pipeline. Both the bulge and disk
components exhibit increased scatter relative to the scatter of the global
parameters. In addition, the bulge component exhibits a systematic
underestimation of the radius, S{\'e}rsic index, and magnitude, particularly for
bulges with larger radii or higher S{\'e}rsic index. 

The galaxies fit in \catalog\ have a median size roughly equivalent
to the average PSF of SDSS. For most galaxies, the
resolution necessary to accurately resolve bulge substructure is not present. As
shown in Section~\ref{subsec:sn_psf}, the ability to recover small bulges is
improved by a factor of 2 increase in resolution. Finer resolution in central
regions of the galaxy is also necessary to fully recover larger S{\'e}rsic
indexes without bias. Even with these systematics, the two-component fits
are still necessary to recover unbiased global parameters and can
provide insight  into the structure of galaxies.

The use of two-component models is potentially ill-advised for many
SDSS galaxies as the respective sub-components may be too small to be well-resolved.
This is suggested by \cite{simard11} as well as \cite{Lackner2012} (if we use the
suggested resolution cut based on the PSF FWHM). However, our data show that
this recommendation should be conditional on the galaxy
parameters of interest. While it may be true that bulge parameters of the
\SerExp\ fit become unreliable at small radii, we show that using only the \Ser\ fit
radius will bias a sample of SDSS galaxies containing both single and
two-component profiles (see Figure~\ref{subfig:psf_serexp_ser}). However, there
are no cases where the \SerExp\ fit introduces bias. It is advisable to
use the \SerExp\ halflight radius and magnitude as the total magnitude
of the galaxy when examining a sample such as this.

 The F-test offers a potentially powerful way to
distinguish when it is necessary to use a more complicated two-component model. The
F-test can compare the $\chi^2$ values
among nested linear models with Gaussian errors \citep[][]{LuptonStat}. Although our
models are not linear and our error distribution is not strictly Gaussian, we  
apply the F-test to our fits. Following \cite{simard11}, we adopt an F-test
probability of 0.32 as the cutoff indicating a more complicated model is required.
When we find a low F-test probability, P$_{correct}<0.32$, the more
complicated model (\ie going from a one-component to two-component fit, or allowing
the S{\'e}rsic index of the bulge to vary) provides a better fit to the
observed profile. In cases where a \Ser\ fit is used rather than a \SerExp\
fit, the improvement in fitting is large enough to justify using a model
with more free parameters. The improved fit is not merely the result of
using a more flexible model. A similar test was performed by \cite{Lackner2012}
to select among a pure disk or disk+bulge model. 

If the selection based on the F-test is correct, then the
resulting measurements of total magnitude and halflight radius will be unbiased.
Using the \SerExp\ mocks fit with each of the \Ser\ and \SerExp\ models, we
select the
fitted model by performing the F-test comparing the \Ser\ and \SerExp\ fits.
The preferred fit (either \Ser\ or \SerExp) of the \SerExp\
mocks is then used to assess the bias in the halflight radius and magnitude.

\begin{figure*}    
\includegraphics[width
=0.30\linewidth]{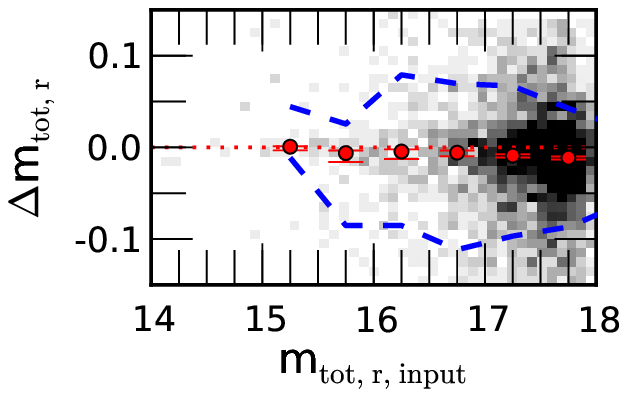}
\includegraphics[width =
0.30\linewidth]{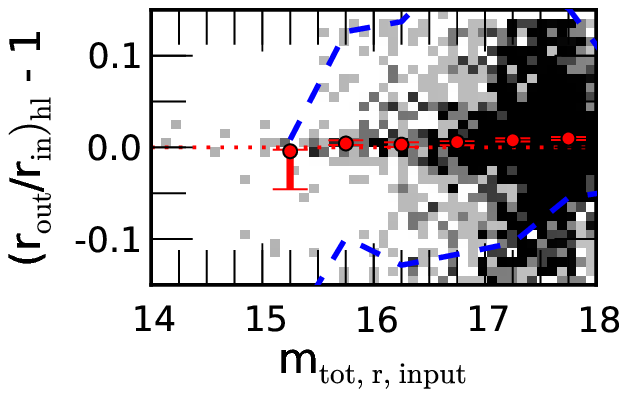}
\includegraphics[width =
0.30\linewidth]{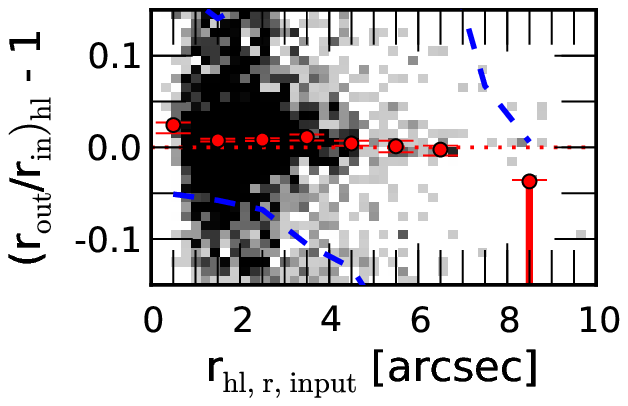}
\caption{Total magnitude and halflight radius of \Ser\ fits of \SerExp\ mocks
shown by F-test to be sufficiently well fit by \Ser\ models. The fits are unbiased,
but the scatter in the recovered values are approximately twice as wide in halflight
radius and magnitude as compared to the \SerExp\ fits in
Figure~\ref{subfig:psf_serexp_serexp}.}
\label{fig:ftest}
\end{figure*}

By examining the subset of \SerExp\ mocks for which the F-test
determines the \Ser\ model to be the appropriate fit, we test the ability of the
F-test to select galaxies that are correctly represented by \Ser\ models. In
Figure~\ref{fig:ftest} we show the resulting distribution of total magnitude and
halflight radius of this subset of \SerExp\ mocks fit with \Ser\ models. The bias 
originally observed in Figure~\ref{subfig:psf_serexp_ser} is not
evident. However, the scatter in the recovered values are approximately twice as
wide as in Figure~\ref{subfig:psf_serexp_serexp}, indicating that while the fits
are unbiased, some sensitivity is lost by using the simpler  (and ultimately
incorrect) model. The remaining \SerExp\ mocks, for which the \SerExp\ fit is
determined by F-test to be most appropriate, are also unbiased in total
magnitude and halflight radius. From this test, we conclude that using the F-test
to determine the most appropriate fitted model allows for unbiased
measurement of the halflight radius and total magnitude. 

Using the \Ser\ mocks, the false positive rate 
(\Ser\ mocks classified as needing a \SerExp\ fit according to the F-test) 
for the F-test with a significance level of 0.32 is 5\%, 
suggesting that there is a low level of contamination in a two-component sample selected using the F-test. 
Using \SerExp\ mocks with $0.2<B/T<0.8$ and $n_{bulge}>2$, which we consider true two-component galaxies,
the false negative rate (\SerExp\ mocks classified as needing only a \Ser\
fit according to the F-test) is 34\%, missing a substantial fraction of the
galaxies with two components. While selection using the F-test is sufficient
to remove the measured bias in global fitting parameters and is able to select a
relatively pure sample of two-component galaxies, it does not select a complete
sample of two-component galaxies. Clearly caution is necessary when using the
F-test to select two-component galaxies from fitting
routines. However, the F-test can indicate when the global parameters of a
\Ser\ model are likely unbiased regardless of the underlying galaxy type.

Following \cite{simard11}, we can also select the fitted model based
on a tiered approach, first performing the F-test on the \Ser\ and \DevExp\ fits.
Galaxies for which the \DevExp\ fit gives a statistically significant improvement
are then tested again to determine whether the \SerExp\ fit is preferable to the
\DevExp\ fit. The preferred fit (either \Ser, \DevExp, or \SerExp) of the \SerExp\
mocks is then used to assess the bias in the halflight radius and magnitude. We
tested this approach and found that it did not significantly alter the results.

Many galaxies exhibit more complex structure than a single- or two-component
structure. Even the case of a two-component model often 
oversimplifies galaxy structure. Bars, 
rings, central sources, clumpyness, or asymmetry cannot be effectively 
modeled in our simulations. Because of this, we can only determine a 
lower-bound on the uncertainty in our parameter estimates. However, 
correcting fits using this lower bound improves the fit of the observed galaxy. 

\begin{figure*}
\includegraphics[width=0.40\linewidth]
{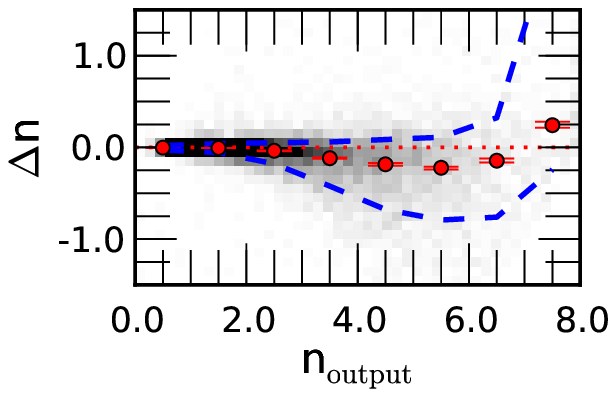}
\includegraphics[width=0.40\linewidth]
{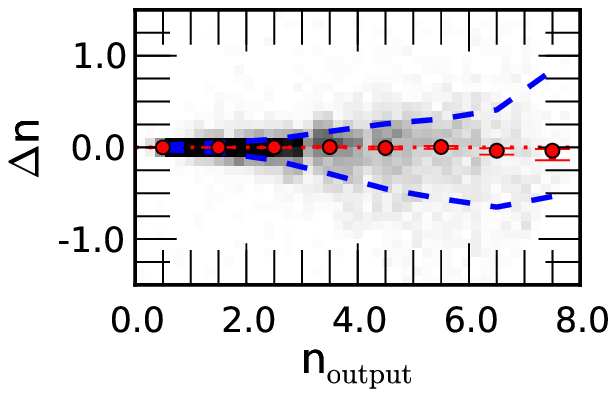}
\includegraphics[width=0.40\linewidth]
{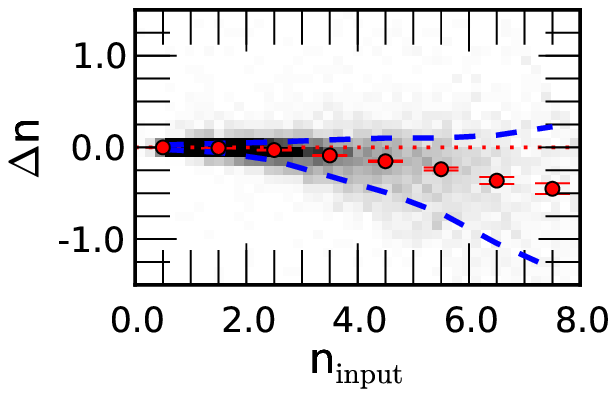}
\includegraphics[width=0.40\linewidth]
{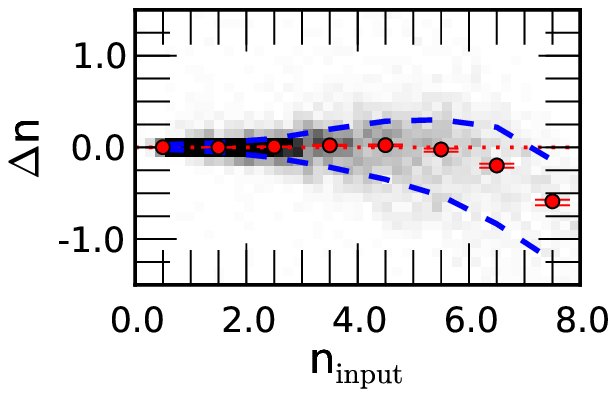}
\caption{An example of the bias correction of the S{\'e}rsic index of the \Ser\
model. The error in the S{\'e}rsic index ($n_{output}$-$n_{input}$) versus
output value is presented
before (top left panel) and after (top right panel) correction. The same
correction is shown in the bottom row versus the simulated value of the
S{\'e}rsic index. We apply the correction in the $n_{output}$ basis. This
appropriately corrects the bias as a function of $n_{input}$ except at high
$n$ where the correction fails. The reason for this failure is
due to the boundaries of the allowed $n$ parameter space. Galaxies in the
highest bins of output S{\'e}rsic index are a combination of poorly fit, low
S{\'e}rsic index galaxies that are artificially constrained to fall in the high
bins, and correctly fit, high S{\'e}rsic index galaxies. The result of this
mixture is a net negative correction on galaxies with high S{\'e}rsic index.}
\label{fig:ser_bias}
\end{figure*}

 We can apply a simple example of bias correction following the
procedure outlined in \cite{Groth2002}. Given the simulated and fitted values
of the S{\'e}rsic index for the \Ser\ model, we plot the bias as a function of
the fitted value output by PyMorph. In this case, the output value represents
the measured value in real data. The simulated value represents the true
underlying value of the galaxy S{\'e}rsic index. We can determine an average
bias and uncertainty in the bias, labeled as $Bias$ and $\Delta Bias$, as a
function of output S{\'e}rsic index. Additionally, we can measure the random
error in the fits from the width of the bias distribution as a function of
S{\'e}rsic index, labeled as $\Delta Random$. Then the corrected S{\'e}rsic
index and uncertainty on the corrected index is
\begin{equation}\label{eq:corrected}
\begin{aligned}
& n_{corrected}  = n_{fitted} - Bias(n_{fitted}) \\
& \Delta n   = \sqrt{\Delta_{galfit}^2 + \Delta Bias^2 + \Delta Random^2} \\
\end{aligned}
\end{equation}
Applying this correction allows us to correct bias as a function of
both simulated and fitted S{\'e}rsic index for the sample of galaxies used in
\catalog. We show the results of this process
in Figure~\ref{fig:ser_bias}.

We are able to statistically correct for the bias in our sample in
both the simulated and fitted bases for most values of the S{\'e}rsic index.
However, there is an under-correction at high simulated S{\'e}rsic value. This
effect appears to be due to the boundaries of the parameter space that PyMorph
is allowed to search for the best fit model. By restricting PyMorph to values of
$n<8$, galaxies simulated with S{\'e}rsic index of 8 will be preferentially
underestimated. However, the highest bins of fitted S{\'e}rsic index
contain many more galaxies with over-estimated S{\'e}rsic index. Therefore the
net correction will be negative and not appropriate for the highest
bins. We could improve the error correction at higher bins by allowing GALFIT to
explore larger values of the S{\'e}rsic index. However, this is beyond the scope of
this paper. 

Additional corrections may also be considered (\ie divide in both magnitude and
S{\'e}rsic index prior to computing the bias correction) depending on the
specifics of a given study. For properties of the global population,
the corrections measured in this paper are applicable to the sample presented in
\catalog.

Our tests were performed on r-band data from SDSS. The performance of the
pipeline can change when observing in different bands. This change is primarily
dependent on the change in the S/N and resolution between bands (due to the
changing brightness of the sky, color of the galaxy, and size
relative to the PSF)
and on the different galactic structures to which neighboring SDSS filter bands
are sensitive. In principle, these effects could be measured from
the simulations presented in this paper by adjusting the S/N and background
level. Additionally, one may have to adjust the size of the galaxies or redraw
the sample to match the size distributions in the different band.  
In \catalog, we fit the SDSS g, r, and i band data. 
It is unlikely that the images change drastically enough over the wavelength 
and redshift range observed to require additional testing in the i 
band. However, these simulations become an increasingly poor estimate 
of error in bluer bands where the photometry becomes more sensitive to 
star forming regions. These regions tend to be clumpier and, therefore, 
less well represented by a smooth profile. Therefore, g band fits may 
present more scatter than the r or i band data. These clumpy regions are
difficult to model with the smooth models presented here. One could
attempt a hybrid approach to generating simulated data whereby one
isolates clumpyness in nearby galaxies and use this as a template to add
clumpyness to smaller SDSS galaxies. However, the details of this process are
beyond the scope of this paper.  

It is also potentially useful to use information about the r-band to inform the
fits of neighboring bands. Indeed \cite{simard11} attempted this by requiring
many parameters (\ie S{\'e}rsic index, radius, ellipticity) of the fitting
model to be identical across the g and r bands, essentially using the two bands
as a form of coadded data to increase the S/N. This increase of S/N comes at
the expense of dis-allowing variation in the matched parameters, which may or
may not be an appropriate assumption (\ie in a two-component fit, we might
expect the bulge size to change across bands, which is dis-allowed).
Additionally, \cite{Haussler2012} enforced simple polynomial relationships in
parameters across bands, using the neighboring bands to further constrain the
acceptable parameter space to be searched by the fitting algorithm. The
most flexible method is to fit each band independently and examine the
systematic effects of each band as necessary, making additional
cross-band comparisons including color \citep[for example, see ][]{Lackner2012}.
This is our preferred method for the data presented here and in \catalog.

\section{Conclusion} \label{sec:conclusions}
We presented the simulations used to test fitting of SDSS galaxies
using PyMorph. Simulations of the \Ser\ and \SerExp\ models were 
presented and examined in many different cases. The simulations were
generated using the results of the fits presented in \catalog. We showed
that our simulations are recoverable in the case of no noise, which 
demonstrates that our simulations are correct. We then
showed that we can
recover the parameters in the case of a simulated background and noise 
representative of the average SDSS image (see Figures~\ref{fig:ser_ser},
\ref{fig:ser_serexp},\ref{fig:serexp_ser}, \ref{fig:serexp_serexp}, and
\ref{fig:serexp_serexp_2com}). 

Several individual effects on the fitting were examined. We showed that our
choice of 20 halflight radii for cutout size does not significantly bias our
fitting results (see Figure~\ref{fig:cutout_size_all}). In addition, we examined the
effect of incorrect background
estimation, which can significantly affect fitting results
(Figure~\ref{fig:bkrd_ser_ser} and \ref{fig:bkrd_serexp_serexp}). Effects of
increasing the S/N are somewhat limited for this sample. However, an
increase in the resolution of the sample would greatly improve parameter
measurements, removing many biases in the two-component fits and improving the
estimation radius and S{\'e}rsic index for \Ser\ galaxies as shown in
Figures~\ref{subfig:pix2_ser_ser} and \ref{subfig:pix2_serexp_serexp_2com}.

We also examined the bias created when fitting incorrect models to galaxies.
Fitting a two-component S{\'e}rsic  + Exponential model to what is really 
just a single S{\'e}rsic  results in a noisier recovery of the input parameters, 
but these are not biased (see Figure~\ref{subfig:psf_ser_serexp}); fitting a single
S{\'e}rsic to what is truly a 
two-component system results in an overestimate of 0.05 magnitudes in total magnitude
and 5\% halflight radius for dim galaxies, increasing to 0.1 magnitudes and 10\% for
galaxies at the brighter end of the apparent magnitude distribution (see
Figure~\ref{subfig:psf_serexp_ser}). These biases are used to correct the
systematics of our fitted SDSS sample
and suggest that magnitude and radius values of a \SerExp\ fit are
the least likely to be biased across many galaxy types. Therefore it is
advisable to use \SerExp\ values when examining global parameters for galaxies.

These simulations can be analyzed together with the fits presented in
\catalog\ to give a more detailed understanding of galaxy structure and
formation as presented in \cite{Bernardi2013}.  

\section*{Acknowledgments}
The authors would like to thank the anonymous referee for many useful comments that
helped to greatly improve the paper. AM and VV would also like to
thank Mike Jarvis and Joseph Clampitt for many helpful discussions.

This work was supported in part by NASA grant ADP/NNX09AD02G
and NSF/0908242.

Funding for the SDSS and SDSS-II has been provided by the Alfred P. Sloan
Foundation, the Participating Institutions, the National Science Foundation, the U.S.
Department of Energy, the National Aeronautics and Space Administration, the Japanese
Monbukagakusho, the Max Planck Society, and the Higher Education Funding Council for
England. The SDSS Web Site is http://www.sdss.org/.

The SDSS is managed by the Astrophysical Research Consortium for the
Participating Institutions. The Participating Institutions are the American Museum of
Natural History, Astrophysical Institute Potsdam, University of Basel, University of
Cambridge, Case Western Reserve University, University of Chicago, Drexel University,
Fermilab, the Institute for Advanced Study, the Japan Participation Group, Johns
Hopkins University, the Joint Institute for Nuclear Astrophysics, the Kavli Institute
for Particle Astrophysics and Cosmology, the Korean Scientist Group, the Chinese
Academy of Sciences (LAMOST), Los Alamos National Laboratory, the
Max-Planck-Institute for Astronomy (MPIA), the Max-Planck-Institute for Astrophysics
(MPA), New Mexico State University, Ohio State University, University of Pittsburgh,
University of Portsmouth, Princeton University, the United States Naval Observatory,
and the University of Washington.

\bibliography{bibliography-arxiv}

\end{document}